\makeatletter
\newif\if@francais
\@francaisfalse
\makeatother
\documentclass[a4paper,11pt]{article}

\usepackage[utf8]{inputenc} 
\usepackage{xspace}	
\usepackage[english]{babel}
\usepackage[T1]{fontenc}
\usepackage{lmodern}
\usepackage[margin=24mm,includeheadfoot]{geometry}
\usepackage{wrapfig}
\usepackage{braket}
\usepackage{minitoc}
\usepackage{cite}
\usepackage{graphicx,xcolor} 
\usepackage{soul}
\usepackage{multirow}
\usepackage{array}

\usepackage{float}
\usepackage{pdfpages} 
\usepackage{amsmath, mathtools}
\mathtoolsset{}
\usepackage{amssymb,amsfonts}  
\usepackage{bm,bbm}         
\usepackage{upgreek}           
\usepackage[e]{esvect}         
\usepackage{esint}		      
\usepackage{etoolbox}          
\usepackage{calc}              
\usepackage{icomma}		       
\usepackage{comment}
\usepackage{caption}
\usepackage{subcaption}
\usepackage[titletoc,title]{appendix}

\usepackage{titlesec}

\titleformat{\section}
{\normalfont\Large\bfseries}{\thesection.}{1em}{}

\titleformat{name=\section,numberless}
{\normalfont\Large\bfseries}{}{0em}{}

\AddToHook{env/appendices/begin}{%
	\titleformat{\section}
	{\normalfont\Large\bfseries}{\sectionname~\thesection.}{1em}{}
}

\usepackage[
			]{hyperref} %

\usepackage{float}

\usepackage{braket}
\usepackage{ulem}
\usepackage{xcolor}
\usepackage[export]{adjustbox}
\usepackage{overpic}

\usepackage{authblk}

\newcommand{\bbR}{\mathbb{R}}

\title{A framework for continuous superradiant laser operation via sequential transport of atoms}

\author[1,2,*]{Jana El Badawi}
\author[1]{Marion Delehaye}
\author[2]{Bruno Bellomo}
\affil[1]{Université Marie et Louis Pasteur, SUPMICROTECH, CNRS, Institut FEMTO-ST (UMR 6174), F-25000 Besançon, France}
\affil[2]{Université Marie et Louis Pasteur, CNRS, Institut UTINAM (UMR 6213), Observatoire des Sciences de l’Univers THETA, 41 bis avenue de l’Observatoire, F-25010 Besançon, France}
\affil[*]{corresponding author: jana.elbadawi@femto-st.fr}
\date{}

\begin{document}
\sethlcolor{white}

\maketitle

\begin{abstract}

We perform a theoretical study of a continuous superradiant laser supporting its experimental realization at FEMTO-ST using two sequentially-emitting ensembles of $^{171}\mathrm{Yb}$ atoms coupled to the same Fabry-Perot cavity. Using an open quantum system approach, we identify for the simplest case the parameter space where the laser reaches tens of picowatts of power with a sub-millihertz linewidth. Studying the impact of inhomogeneous frequency broadening and variations in atom–cavity coupling on the superradiant emission, we find the laser properties robust with respect to such perturbations, also thanks to the occurrence of synchronization of the atomic dipoles.  
We then consider a two-site configuration, in which atoms in each site are equally coupled to the cavity and have equal detunings, with different values for the two ensembles. We find for balanced and imbalanced atom numbers that synchronization leads in a certain parameter space to a single narrow spectral line whose central frequency  follows the weighted average frequency. This result indicates that sequential loading can enable continuous superradiant emission for metrological applications, provided that the relative frequencies of the two ensembles are controlled to the level required by the target stability.
\end{abstract}

\section{Introduction}
The field of time and frequency metrology has seen remarkable advancements over recent decades. These developments have paved the way for a wide range of applications, from investigating variations in fundamental constants~\cite{Safronova_2019} to seismology~\cite{Tanaka2021}. Today's most precise timekeepers consist of the combination of a laser pre-stabilized on an ultra-stable, monolithic Fabry-Perot cavity, providing short-term stabilization down to \(4 \times 10^{-17}\) at one second~\cite{Matei2017}, with an atomic ensemble that ensures long-term stabilization down to \(4.8 \times 10^{-17} (\tau/\mathrm{s})^{-1/2}\)~\cite{Oelker2019, Bothwell2022} and accuracy that can reach the \(10^{-19}\) range~\cite{Aeppli2024, Hausser_2025}.
Achieving even lower instabilities in optical clocks could unlock new applications, such as detecting gravitational waves~\cite{Kolkowitz2016} or searching for dark matter~\cite{Derevianko_2016, Filzinger2023}. However, these systems are currently limited by the fundamental thermal Brownian noise of the Fabry-Perot resonators. To overcome this limitation, several projects have emerged worldwide, including spectral hole burning~\cite{Thorpe2011}, Ramsey-Bordé interferometers~\cite{Olson2019}, and superradiant lasers~\cite{Meiser2009}. This article focuses on the project of a superradiant laser.

The concept of a superradiant laser is based on superradiance, first introduced by Dicke in 1953~\cite{Dicke1953}, and further explored in the context of lasers in~\cite{Gross1982, Haake1993}. About 15 years ago, the idea of using such a laser as a new frequency reference emerged, with the potential to outperform current optical clocks by an order of magnitude~\cite{Meiser2009}. This idea has generated significant excitement, leading to several superradiant laser experiments aimed at metrological applications worldwide~\cite{Bohnet2012, Norcia2016, Norcia2016a, Norcia2017, Laske2019, Kristensen2023, Bohr_2024}.

To date, all reported experiments operate in a pulsed manner, with the emission duration of one pulse below 1 second. Promising stabilities in the \(10^{-16}\) range have been reported~\cite{Norcia2017}, and sub-natural linewidth emission has been observed~\cite{Kristensen2023}. However, the current pulsed nature of the emission remains a significant limitation for its use as a frequency reference. Intense theoretical efforts have been devoted to exploring possibilities for achieving continuous emission~\cite{steady_Holland, Kazakov2013, Hotter2022, Dubey2024}, which predict an appealing ultimate frequency instability at the \(10^{-18}\) level at one second integration time~\cite{Kazakov2022}. One currently investigated possibility involves producing a beam of cold atoms loaded into a conveyor belt inside a bow-tie cavity where superradiance occurs~\cite{Dubey2024}. Other experiments focus on hot beams of atoms crossing a Fabry-Perot cavity to exploit the sub-natural linewidth emission achievable with such a system~\cite{Liu2020, Tang2022, Laburthe-Tolra2023, Oh2024}.

Here, we theoretically investigate a system in which continuous superradiant emission could be achieved using two sequentially-emitting ensembles of ${}^{171}\mathrm{Yb}$ atoms coupled to the same Fabry-Perot cavity, in order to realize predictions for the experiment under development at FEMTO-ST described in \autoref{sec:femto}. We use the framework of open quantum systems and the second-order cumulant expansion to describe the system, and we use the quantum regression theorem to obtain the emission spectrum in \autoref{sec:model}.

In the first description of the superradiant laser system, a fixed number of identical $^{87}$Sr atoms was considered \cite{Meiser2009}. This description captured the fundamental physics of the collective atomic dynamics and showed that it can produce a mHz linewidth laser with useful power. In ~\autoref{1class} we adopt a similar approach for our system of $^{171}$Yb atoms and show the characteristics of such a laser. In a recent study, this framework was extended to more realistic experimental conditions, in which the atoms have varying transition frequencies and are considered to be coupled to the cavity with varying strengths \cite{Bychek_2021}. 
In ~\autoref{Xclass} we follow this approach using experimental parameters relevant to our system, and demonstrate the robustness of the emission with respect to such perturbations.

Following this, we explore in ~\autoref{sec:2sites} the effects of a key feature of FEMTO-ST experiment, the loading of two sites labeled $A$ and $B$ inside the cavity, each loaded by a group of atoms with equal parameters. First, in~\autoref{section:NA=NB}, we revisit the situation explored theoretically in \cite{Minghui_NA=NB} where the total number of atoms in each site is equal. In \cite{Minghui_NA=NB} this is explored by adiabatically eliminating the cavity field while here we keep the full description of the system.
Second, in ~\autoref{section:NAdiffNB}, we consider the realistic experimental implementation for our system where the atoms are going to be transported to each site sequentially and consider that the total number of atoms in each site can vary. The characteristics of the laser is shown with the impact of atom number imbalance between the two sites.

\section{Experimental concept}\label{sec:femto}

At FEMTO-ST, we are establishing a superradiant experiment to explore the spectral capabilities of this new ultra-stable frequency reference. Our approach involves an original scheme to achieve sustained superradiant emission on the \(^{171}\mathrm{Yb}\) \(^1\mathrm{S}_0 \rightarrow {}^3\mathrm{P}_0\) clock transition at 578 nm, which has a natural linewidth of 7 mHz.
Our experimental apparatus consists of two chambers. In the first chamber, we prepare a magneto-optical trap \hl{of up to $10^8$} \(^{171}\mathrm{Yb}\) atoms, \hl{and we expect to transport up to $10^6$ atoms into one zone of a 56 mm-long Fabry-Pérot cavity. This defines the practical operating range considered in this work. The cavity has a finesse of approximately $7000$ at the clock transition wavelength and the atoms are coupled to a single transverse cavity mode with an estimated mode diameter of about $145~\text{\textmu m}$.}

\begin{figure}[H]
\centering
\begin{subfigure}[t]{0.31\textwidth}
    \begin{overpic}[width=\textwidth, valign=t, trim={15cm 1cm 15cm 5cm}, clip]{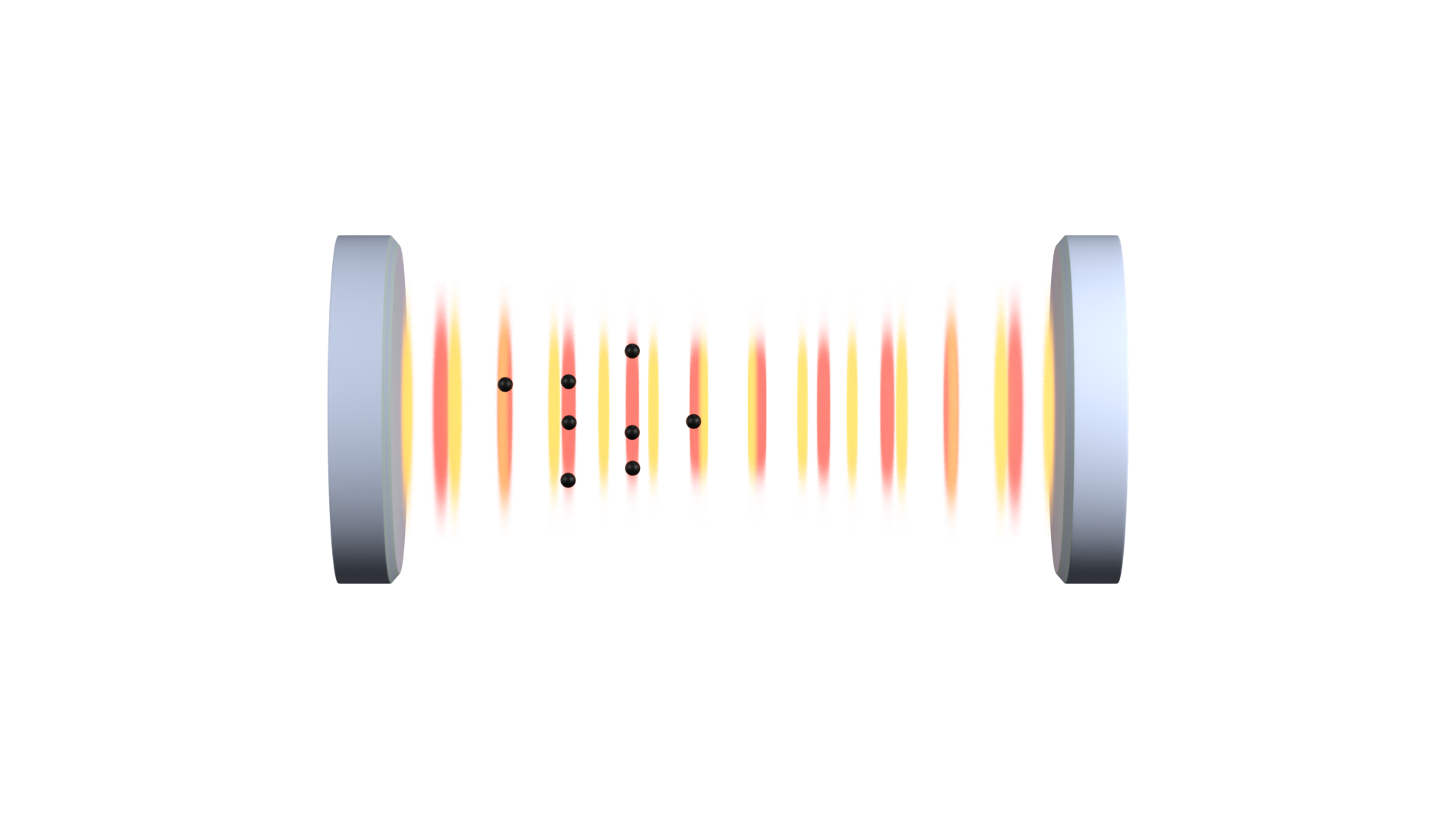}
        \put(0,74){\textbf{(a)}}
        \put(31,70){$A$}
        \put(62,70){$B$}
    \end{overpic}
\end{subfigure}
\hfill 
\begin{subfigure}{0.31\textwidth}
   \begin{overpic}[width=\textwidth, valign=t, trim={15cm 1cm 15cm 5cm}, clip]{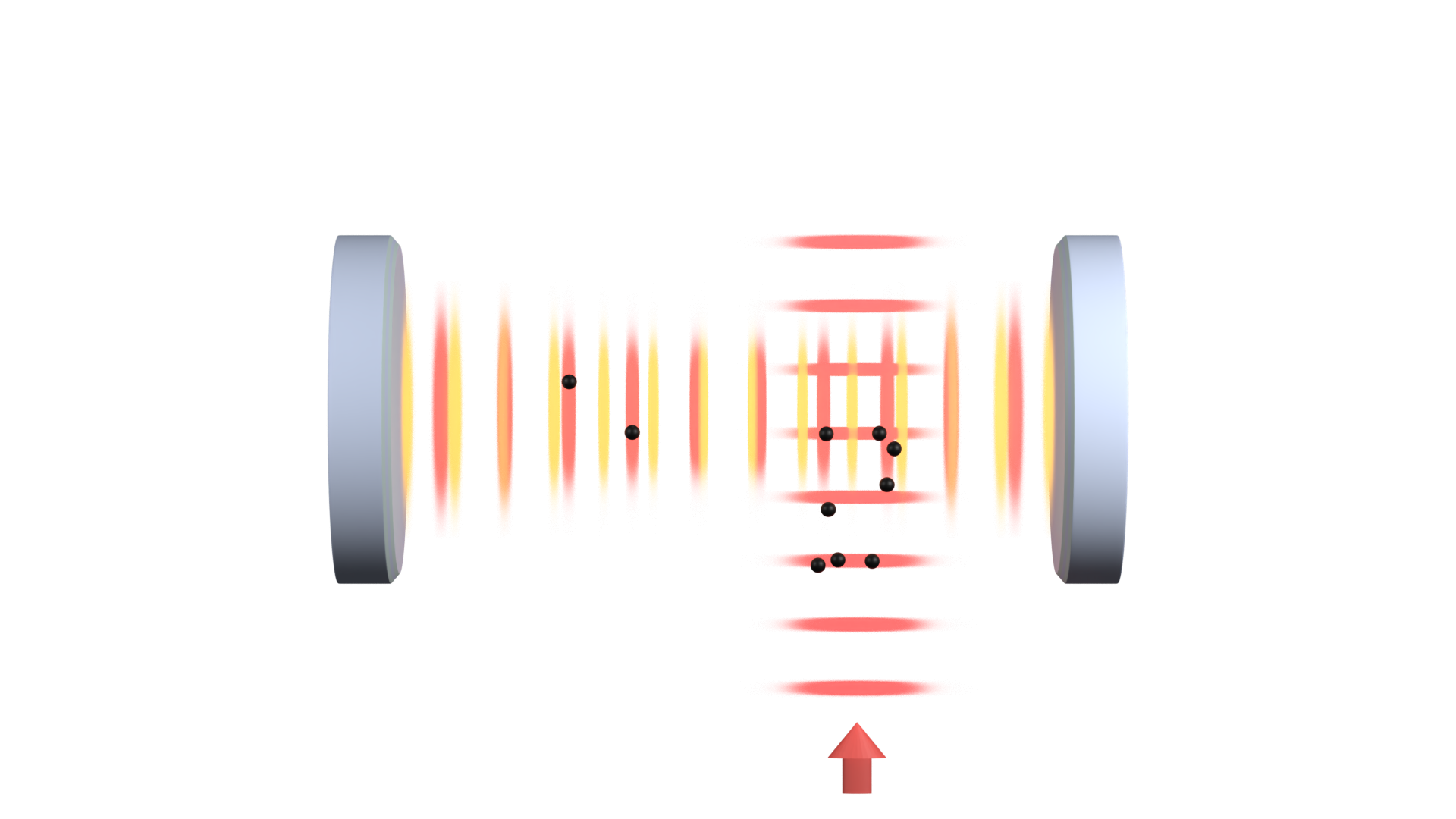}
          \put(0,74){\textbf{(b)}}
    \end{overpic}
\end{subfigure}
\hfill
\begin{subfigure}{0.31\textwidth}
    \begin{overpic}[width=\textwidth, valign=t, trim={15cm 1cm 15cm 5cm}, clip]{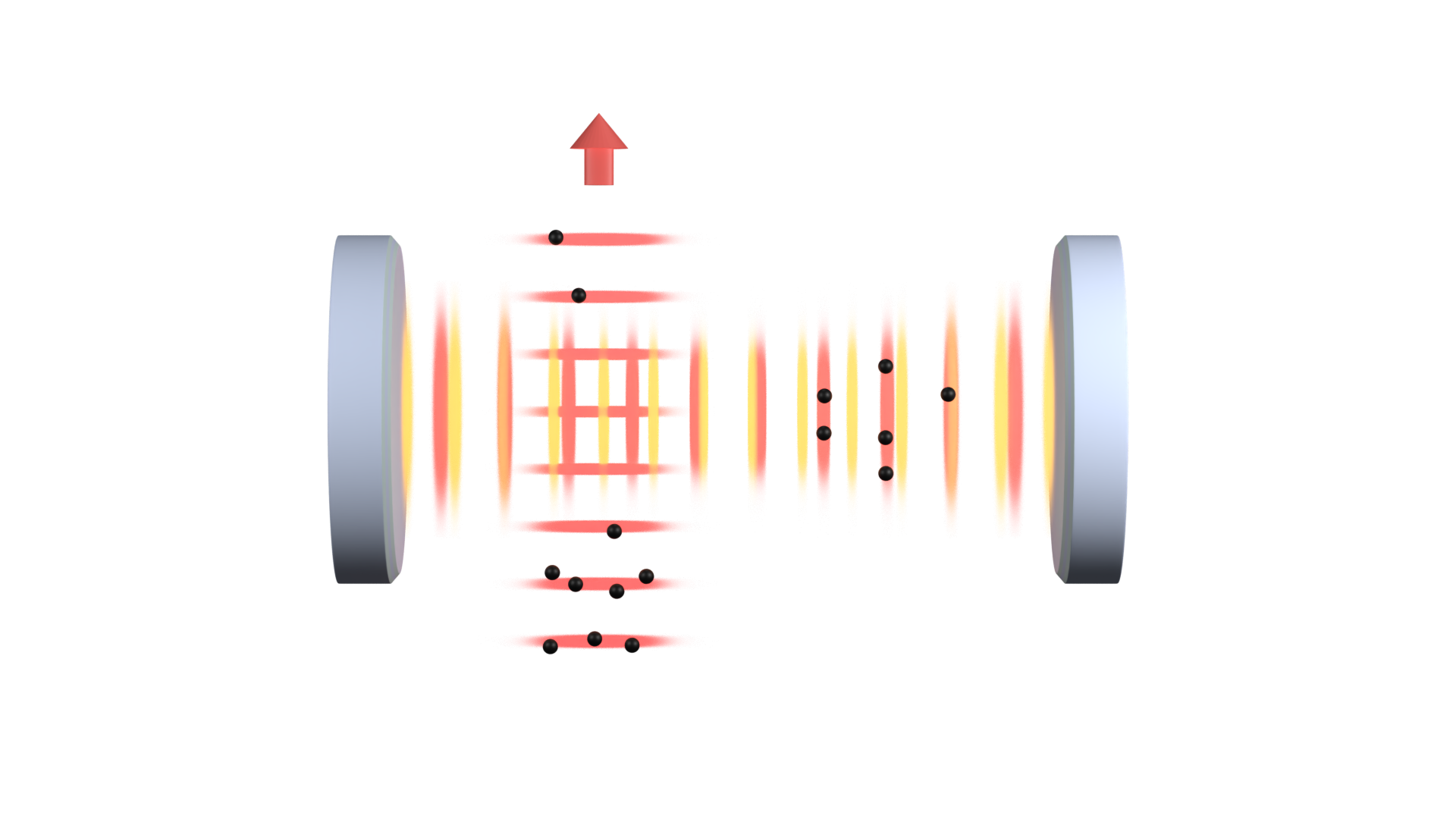}
        \put(0,74){\textbf{(c)}}
    \end{overpic}
\end{subfigure}
\caption{
Scheme of the experiment at FEMTO-ST. Black dots: atoms. Yellow lattice: cavity mode. Red lattices: trapping lattice within the cavity and optical conveyor belts. (a) Atoms are trapped in site $A$ of the cavity. (b) The number of atoms in site $A$ has decayed, hence a new ensemble of atoms is loaded into the site $B$ of the cavity by an optical conveyor belt. (c) Atoms are trapped in site $B$, and a new ensemble of atoms is prepared to be loaded in site $A$.
}
\label{fig:femto_exp}
\end{figure}

\begin{figure}[b!]
    \centering
    \includegraphics[width=0.3\linewidth]{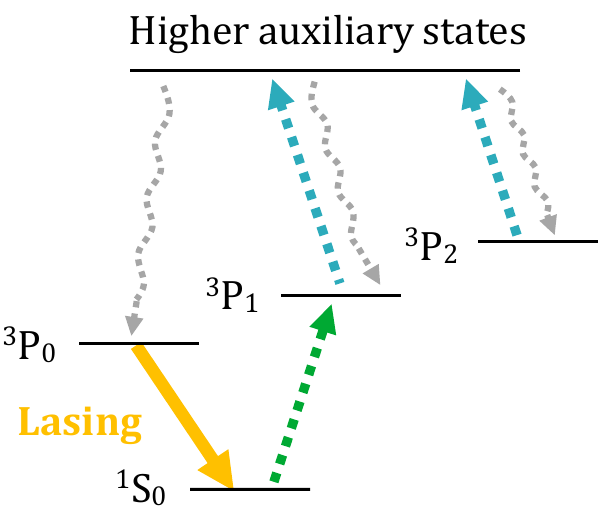}
    \caption{\hl{Proposed repumping scheme for the continuous superradiant laser based on ${}^{171}\mathrm{Yb}$ atoms. The superradiant lasing occurs on the clock transition ${}^1\text{S}_0\leftrightarrow {}^3\text{P}_0$. The intermediate states ${}^3\text{P}_1$ and ${}^3\text{P}_2$ participate in the repumping cycle which involves also  higher auxiliary states, as shown schematically in the figure.}
    }
    \label{fig:repumping-scheme}
\end{figure}

\hl{The experiment is based on two fixed zones in the cavity, referred to as ``site $A$'' and ``site $B$'', separated by approximately 3~mm. The two sites are sequentially populated using an optical conveyor belt in order to maintain the superradiant emission, as shown in} Figure~\ref{fig:femto_exp}.
\hl{In a typical experimental sequence, an atomic ensemble is first loaded into site $A$. This ensemble will be continuously repumped via a multi-step process which involves several auxiliary levels, as shown qualitatively in} Figure~\ref{fig:repumping-scheme}\hl{, similarly to what is described in}~\cite{Hotter2022}\hl{. This mechanism gives rise to an effective repumping rate used in the model described in section}~\ref{sec:model}.
As the atom number decays over time due to residual heating and collisions with the background gas, we will prepare another ensemble to be loaded into site $B$ before the emission from site $A$ ends \hl{with a targeted refresh rate of $1$-$2$~Hz}. In this paper, we investigate theoretically the spectral properties of the superradiant light emitted by such a system.

\hl{${}^{171}\mathrm{Yb}$ has a nuclear spin $\mathrm{I}=1/2$, so both the ground state~${}^1\mathrm{S}_0$ and the excited clock state~${}^3\mathrm{P}_0$ have two Zeeman sublevels. An external magnetic field will be applied orthogonally to the cavity axis, which will shift the sublevels symmetrically around the bare atomic frequency. The average of the emissions from the Zeeman sublevels allows for the reconstruction of a signal that is insensitive to the magnetic field}~\cite{Norcia2017}. For typical magnetic fields of a few gauss, the resulting frequency shifts are on the order of a few hundreds Hz, which is larger than the targeted repumping rate, as found later in section~\ref{1class}. In this paper, we consider a single pair of lasing states.

\section{Description of the model}\label{sec:model}

As model for the experimental setup described in the previous section and shown schematically in Figure~\ref{fig:femto_exp},
we consider an ensemble of $N$ cold $^{171}\mathrm{Yb}$ atoms confined in an optical lattice and treated as two-level systems. Each atom in the ensemble is labeled by an index $i$, and is coupled to the cavity field with a position-dependent coupling strength $g_i$. The atomic transition frequency of the $i$-th atom is denoted by $\omega_i$ and the resonance frequency of the cavity by $\omega_c$. In particular, in the absence of individual atomic shifts, all the atoms are characterized by the unperturbed clock transition frequency $\omega_a$, and the cavity is assumed to be tuned in resonance with this transition. The unitary dynamics of the atoms and the cavity is then governed by the Hamiltonian (if not specified otherwise, we set $\hbar = 1$) 

\begin{equation}
	H=\sum_{i=1}^N\omega_i \sigma_i^{+} \sigma_i^{-}+ \omega_c a^{\dagger}a +\sum_{i=1}^N g_i\left(a \sigma_i^{+}+a^{\dagger} \sigma_i^{-}\right)\!,  
\end{equation}
where $ \sigma_i^{+} = \ket{e}_i\bra{g}_i $ and $ \sigma_i^{-} = \left( \sigma_i^{+} \right)^{\dagger} $ are the raising and lowering operators for the $ i $-th atom, $ \ket{g} $ and $ \ket{e} $ correspond to the atomic ground and excited states, ${}^1\mathrm{S}_0 $ and $ {}^3\mathrm{P}_0 $, respectively, and $ a^{\dagger} $ and $ a $ are the photon creation and annihilation operators of the cavity mode. 

The atoms and cavity mode are subject to various dissipative processes. We consider a zone of the parameters in which the global dynamics for the atoms-cavity density matrix $\rho$ is governed by a master equation in the Lindblad form,
\begin{equation}
	\frac{d}{dt}\rho=-i[H, \rho]+\mathcal{L}[\rho],
\end{equation}
where $\mathcal{L}[\rho]$ contains all the dissipative processes acting on the system via the operator $O$ at a rate $x$, described by the Lindblad superoperator $x \mathcal{D}[O] \rho= x\left(2 O \rho O^{\dagger}-O^{\dagger} O \rho-\rho O^{\dagger} O\right)/2$. The dissipation includes the photons losses from the cavity at a rate $\kappa$, the spontaneous atomic decay at a single-atom rate $\gamma$, the individual incoherent repumping with a rate $R$, and inhomogeneous dephasing of each atom at a rate $1/T_2$. A cavity dephasing term denoted by the parameter $\xi$ which accounts for the effective noise induced by thermal fluctuations of the cavity mirrors is also included as in \cite{Bychek_2021, Kazakov2022}. The single-atom rates $\gamma$, $R$, and $1/T_2$ are assumed to be identical for all atoms.  These various dissipative processes contribute to the following total Lindblad dissipator,

\begin{equation}
	\mathcal{L}[\rho] = \kappa\mathcal{D}[a] \rho + \xi \mathcal{D}[a^{\dagger} a] \rho + \sum_{i=1}^N \left[ \gamma \mathcal{D}\left[\sigma_i^{-}\right] \rho + R \mathcal{D}\left[\sigma_i^{+}\right] \rho + \frac{1}{2 T_2} \mathcal{D}\left[\sigma_i^{+} \sigma_i^{-}\right] \rho\right]\!.
\end{equation}
The zone of the parameters we work in justifies the use of a local approach to derive the total dissipator. In this approach, $\mathcal{L}[\rho]$ is obtained by the sum of the individual contributions from each dissipative source, as if it was the only source acting on the system \cite{breuer2002theory, Rivas_2010, Bellomo2020}.

Superradiant lasers operate in the so-called ``bad-cavity limit''~ \cite{Meiser2009, BonifacioQST1}, in which the cavity decay rate $\kappa$ is much larger than the natural atomic decay rate $\gamma$, i.e., $\kappa\gg\gamma$. For our system the decay rate of the cavity is $\kappa = 2\pi \times 400 ~\mathrm{kHz} \approx 2.5 \times 10^6 ~\mathrm{rad.s^{-1}}$ which is many orders of magnitude more than the atomic decay rate $\gamma = 2\pi \times 7~\mathrm{mHz} \approx 4.4 \times 10 ^{-2} ~\mathrm{rad.s^{-1}}$, corresponding to the ultra-narrow linewidth of the clock transition in the fermionic isotope of ytterbium $^{171}\mathrm{Yb}$. 
The rate of the atom-cavity coupling for our system is $g = 2\pi \times 3.8~\mathrm{Hz} \approx 24 ~\mathrm{ rad.s^{-1}}$, so we have $\kappa\gg g$ in our setup. The $1/T_2$ value of our experiment is not measured yet, so in the following we consider $1/T_2 \approx 1~\mathrm{rad.s^{-1}}$ to be a good approximation~\cite{Boyd_T2}. 

The parameters described above are fixed for our system, as they are determined by the experimental setup and physical characteristics of the components. A parameter that can be experimentally scanned over some range is that of the incoherent repumping rate $R$. This parameter describes the average rate at which atoms are repumped from the ground state ${}^1\mathrm{S}_0$ to the excited state ${}^3\mathrm{P}_0$ of the clock transition. It provides the population inversion required for continuous superradiance. $R$ enters the model as simple rate parameter, however it is an effective description of a complex multi-level repumping scheme within the full atomic structure of $^{171}\mathrm{Yb}$. The value of $R$ depends on the specific repumping configuration and can be calculated using the numerical method presented in \cite{Hotter2022}, which models continuous multi-step repumping for the clock transition in bosonic $^{88}\mathrm{Sr}$. In \cite{Hotter2022}, it is stated that this method can extend well to the fermionic $^{87}\mathrm{Sr}$ and to other alkaline-earth atoms, and that includes $^{171}\mathrm{Yb}$ which is used in our system. As a result, the value of $R$ can have a tunable range based on the configuration of the multi-level system. The rate $\xi$ is associated in the model to a dephasing process for the cavity field, an additional dissipation term for the cavity other than the one governed by the photon loss rate $\kappa$. We follow the approach used in \cite{Bychek_2021} and treat $\xi$ as a tunable parameter to check its impact on the characteristics of the laser. 
In all of this study, $\kappa$ stays the largest decay rate of the system.

\subsection{Second-order cumulant expansion}\label{subsec:2ndcumulant}

As the Hilbert space describing such a system grows exponentially with the atom number $N$, solving the Lindblad master equation becomes computationally complex for large $N$. In such cases, performing appropriate approximations is essential to simplify the problem. From the various approximation methods, the second-order cumulant expansion is commonly used \cite{kubo_Ryogo}. 

As the $N$ atoms interact collectively with the cavity field, high-order atom correlations develop during the dynamics. Accurately considering such correlations is important for describing cooperative phenomena such as superradiant emission. However, accounting for all high-order correlations in a system of many atoms quickly becomes intractable, both computationally and analytically. The second-order cumulant expansion is a powerful approximation method that gives higher-order correlations in terms of lower-order ones.

The key idea behind the second-order cumulant expansion is to truncate the correlations at second order by assuming that third- and higher-order cumulants are negligible. This assumption is valid in many physical systems where three-body, or higher, correlations are weak or can be captured by one- and two-body terms.  The cumulants of arbitrary operators $X_1$, $X_2$, $X_3$ up to third order are defined as:
\begin{equation}
	\begin{aligned}
		\braket{X_1}_c &= \braket{X_1}, \qquad
		\braket{X_1 X_2}_c = \braket{X_1 X_2} - \braket{X_1} \braket{X_2}, \\
		\braket{X_1 X_2 X_3}_c &= \braket{X_1 X_2 X_3}
		- \braket{X_1 X_2} \braket{X_3}
		- \braket{X_1 X_3} \braket{X_2}
		- \braket{X_1} \braket{X_2 X_3}
		+ 2 \braket{X_1} \braket{X_2} \braket{X_3}\!.
	\end{aligned}
\end{equation}
In this framework, the approximation $\braket{X_1 X_2 X_3}_c = 0$ allows us to express third-order correlations approximately as:
\begin{equation}
	\braket{X_1 X_2 X_3}_c = 0 \Longrightarrow 
	\braket{X_1 X_2 X_3} \approx 
	\braket{X_1 X_2}\braket{X_3} + 
	\braket{X_1 X_3}\braket{X_2} + 
	\braket{X_1}\braket{X_2 X_3} - 
	2\braket{X_1}\braket{X_2}\braket{X_3}\!.
\end{equation}

This second-order cumulant expansion greatly reduces the computational cost while conserving the two-body quantum correlations. In the context of the linewidth of a superradiant laser, it was shown that this method, when compared with the full quantum solution, provides more accurate results than the mean-field and Langevin-based methods \cite{PhD_SwadheenDubey}.

The total phase invariance of the system is conserved under the second-order cumulant expansion and, as a result, the phase-dependent terms such as $\braket{a}$, $\braket{a^{\dagger}}$, and $\braket{\sigma^{\pm}}$ are zero \cite{Meiser2009}. Starting from the master equation with the Hamiltonian of the system in the rotating frame of the cavity $H_{r} = -\sum_{i=1}^N\Delta_i\sigma_i^+\sigma_i^- +\sum_{i=1}^Ng_i(a\sigma_i^+ + a^{\dagger}\sigma_i^-)$, where $\Delta_i = \omega_c - \omega_i$ is the detuning between the $i$-th atom and the cavity, and applying the second-order cumulant expansion, we obtain the following closed set of equations that describe the system’s dynamics:
\begin{equation}\label{eqs:Steady_state}
	\begin{array}{lll}
		\displaystyle \frac{d}{dt} \braket{a^{\dagger} a} 
		& = & \displaystyle 
		-\kappa \braket{a^{\dagger} a} 
		+ \mathrm{i} \sum_{i=1}^N g_i \left(\braket{a \sigma_i^{+}} 
		-  \braket{a^{\dagger} \sigma_i^{-}}\right) ,\\
		
		\displaystyle \frac{d}{dt} \braket{a \sigma_i^{+}} 
		& = & \displaystyle 
		- \frac{1}{2}\left(\kappa + \gamma + R + \xi + \frac{1}{T_2} + \mathrm{i} 2\Delta_i \right) \braket{a \sigma_i^{+}} 
		+ \mathrm{i} g_i \braket{a^{\dagger} a} 
		- 2\mathrm{i} g_i \braket{a^{\dagger} a} \braket{\sigma_i^{+} \sigma_i^{-}} \\
		&   & \displaystyle 
		- \mathrm{i} g_i \braket{\sigma_i^{+} \sigma_i^{-}} 
		- \mathrm{i} \sum_{j,j\neq i}^N g_j \braket{\sigma_{i}^{+} \sigma_{j}^{-}}, \\
		
		\displaystyle \frac{d}{dt} \braket{\sigma_i^{+} \sigma_i^{-}} 
		& = & \displaystyle 
		\mathrm{i} g_i \braket{a^{\dagger} \sigma_i^{-}} 
		- \mathrm{i} g_i \braket{a \sigma_i^{+}} 
		- (\gamma + R) \braket{\sigma_i^{+} \sigma_i^{-}} + R, \\
		
		\displaystyle \frac{d}{dt} \braket{\sigma_{i}^{+} \sigma_{j}^{-}} 
		& = & \displaystyle 
		-\mathrm{i}(\Delta_i - \Delta_j)\braket{\sigma_{i}^{+} \sigma_{j}^{-}}
		+ \mathrm{i} g_i \braket{a^{\dagger} \sigma_j^{-}} 
		- \mathrm{i} g_j \braket{a \sigma_i^{+}} 
		- 2\mathrm{i} g_i \braket{a^{\dagger} \sigma_j^{-}} \braket{\sigma_i^{+} \sigma_i^{-}} 
		\\
		&   & \displaystyle 
		+ 2\mathrm{i} g_j \braket{a \sigma_i^{+}} \braket{\sigma_j^{+} \sigma_j^{-}}- \left(\gamma + R + \frac{1}{T_2}\right) \braket{\sigma_{i}^{+} \sigma_{j}^{-}}\!.
	\end{array}
\end{equation}

In the following sections, we handle specific cases of this system and we only focus on its steady-state solutions. First, in \autoref{1class}, we consider identical atoms equally coupled to the cavity mode, and this reduces Eqs.~\eqref{eqs:Steady_state} to a set of four equations. Note that in this case there is symmetry in the terms with respect to exchange of atoms. For instance, $\braket{a^{\dagger} \sigma_j^{-}} =\braket{a^{\dagger} \sigma_1^{-}}$ for all $j$, and $ \braket{\sigma_i^{+}\sigma_j^-}=\braket{\sigma_1^{+} \sigma_2^{-}}$ for all $i \neq j$. The analytical steady-state solutions for this case are presented in \autoref{app:ss}, with further simplifications in the large $ \kappa$ and large $ R $ limit in \autoref{app:simple}, and large $ N $ limit in \autoref{app:largeN}. Next in \autoref{Xclass}, we consider the case of non-identical atoms, where detunings can differ, $\Delta_i \neq \Delta_j$.

To manage with the increasing number of equations, we adopt an approach commonly used in the literature \cite{Bychek_2021}, where the atoms are grouped into $M$ frequency classes. Each class contains atoms that share the same detuned frequency, and the number of atoms are distributed according to a Gaussian frequency distribution. Similarly, in the case of atoms with different coupling strengths to the cavity mode, the atoms can be grouped in $K$ coupling strength classes. In the presence of both inhomogeneous broadening and varying coupling strengths, the system is described by $M \times K$ classes. The atoms are divided equally among $K$ coupling classes, and within each coupling class they are distributed into $M$ frequency classes. 

Last, in section~\autoref{sec:2sites} we consider the experimentally relevant scenario with two groups of atoms with their own detuning $\Delta_A$ and $\Delta_B$, and consider the case of equal number of atoms and the case of an imbalance in the number of atoms.

\subsection{Spectrum of the cavity field}\label{sec:spectrum}

Having formulated the dynamics within the second-order cumulant framework, we now focus on the characteristics of the spectrum of the cavity field. To calculate this spectrum, we make use of the Wiener–Khinchin theorem, according to which the spectrum $S(\omega)$ is given by the Fourier transform of the first-order correlation function $g^{(1)}(\tau) = \braket{a^{\dagger}(\tau+t) a(t)}$ \cite{Puri2001}, 
\begin{equation}
	S(\omega)=2 \operatorname{\bbR} \left\{  \int_0^{\infty} \mathrm{d} \tau e^{-\mathrm{i} \omega \tau} g^{(1)}(\tau) \right\}.
\end{equation}
To compute this correlation function, we apply the quantum regression theorem, which states that the time derivative of $ \braket{a^{\dagger}(\tau+t) a(t)}$ follows the time derivative of $\braket{a^{\dagger}(t)}$, and we obtain using the second-order cumulant expansion: 
\begin{equation}
	\begin{aligned}
		\frac{d}{d \tau} \braket{a^{\dagger}(\tau+t) a(t)} & = -\frac{\kappa+\xi}{2} \braket{a^{\dagger}(\tau+t) a(t)} + \mathrm{i} \sum_{i=1}^N g_i \braket{\sigma_i^{+}(\tau +t) a(t)}, \\
		\frac{d}{d \tau } \braket{\sigma_i^{+}(\tau +t) a(t)} & = -\mathrm{i} g_i \braket{\sigma_i^z(\tau + t)} \braket{a^{\dagger}(\tau+t) a(t)} - \frac{1}{2}\left( \gamma+R+ \frac{1}{T_2} + 2 \mathrm{i}\Delta_i \right) \braket{\sigma_i^{+}(\tau +t) a(t)}.
	\end{aligned}
	\label{eq:time_evo}
\end{equation}
To solve the above equations, the starting point $t=0$ is taken when the system is in steady-state. The term $\braket{\sigma_i^z(\tau+t)}$ is then considered a constant equal to the steady-state value $\braket{\sigma_i^z}^{\mathrm{st}}$, which can be obtained from Eqs.~\eqref{eqs:Steady_state}. Analytical expressions for the spectrum are provided in \autoref{app:spectrum} for the one-class case. 

The cavity spectrum provides access to key observables of the system, such as the spectral linewidth and the power and the output of the cavity. The spectral linewidth corresponds to the full-width at half-maximum (FWHM) of the spectrum. The output power, on the other hand, is determined by the mean intracavity photon number $\braket{a^{\dagger}a}$ and is given by $P = \eta\hbar\omega_0\kappa\braket{a^{\dagger}a}$, where $\eta$ is the relative transmission of the output mirror of the cavity and $\omega_0$ is the central angular frequency of the spectrum \cite{Kazakov2022}. In what follows, all power calculations are performed with $\eta=1$, that corresponds to ideal transmission. When all the atoms are on resonance with the cavity field, the spectrum is expected to be centered at $\omega_0=\omega_c$.

\section{One-class case}\label{1class}

In this section we consider the case of identical atoms ($\Delta_i = \Delta_j=\Delta)$ equally coupled to the cavity mode ($g_i = g_j = g$). In particular, we focus on the case when the atoms are on resonance with the cavity mode ($\Delta =  0)$ since the non-resonant case gives similar results if $\Delta$ is small enough. This follows directly from Eqs.~\eqref{Steady state relations}, which imply that for $\Delta\ll (\gamma +\kappa + (1/T_2) +\xi +R)/2$, all the variables of Eqs.~\eqref{eqs:Steady_state}
do not depend appreciably on $\Delta$. We also notice that in this regime $\mathbb{R}\left[\braket{a\sigma^{+}}^\mathrm{st}\right]$, which is zero for $\Delta=0$, is much smaller than $ \mathbb{I}\left[\braket{a\sigma^{+}}^\mathrm{st}\right]$.
The parameters are set equal to the experimental values provided in \autoref{sec:model}, $\gamma = 2\pi \times 7~\mathrm{mHz} $, $\kappa = 2\pi \times 400~\mathrm{kHz} $, $g = 2\pi\times 3.8~\mathrm{Hz}$, and $1/T_2 = 1 ~\mathrm{rad.s^{-1}}$, with $\xi = 0$ except when specified otherwise. In what follows, we study the steady-state solutions of the system described by Eqs.~\eqref{eqs:Steady_state}. The solutions are found numerically in the main text, though for this section they can be computed using the analytical derivations in \autoref{app:ss}. We also observe that according to the derivation of the spectrum for $\Delta=\xi=0$ in Eq.~\eqref{app:Spectrum special case}, when $\braket{\sigma^z}^\mathrm{st}>0$ the spectrum has only one peak and is symmetric with respect to the cavity frequency.

\begin{figure}[b!]
	\centering
	\includegraphics[width=0.7\linewidth]{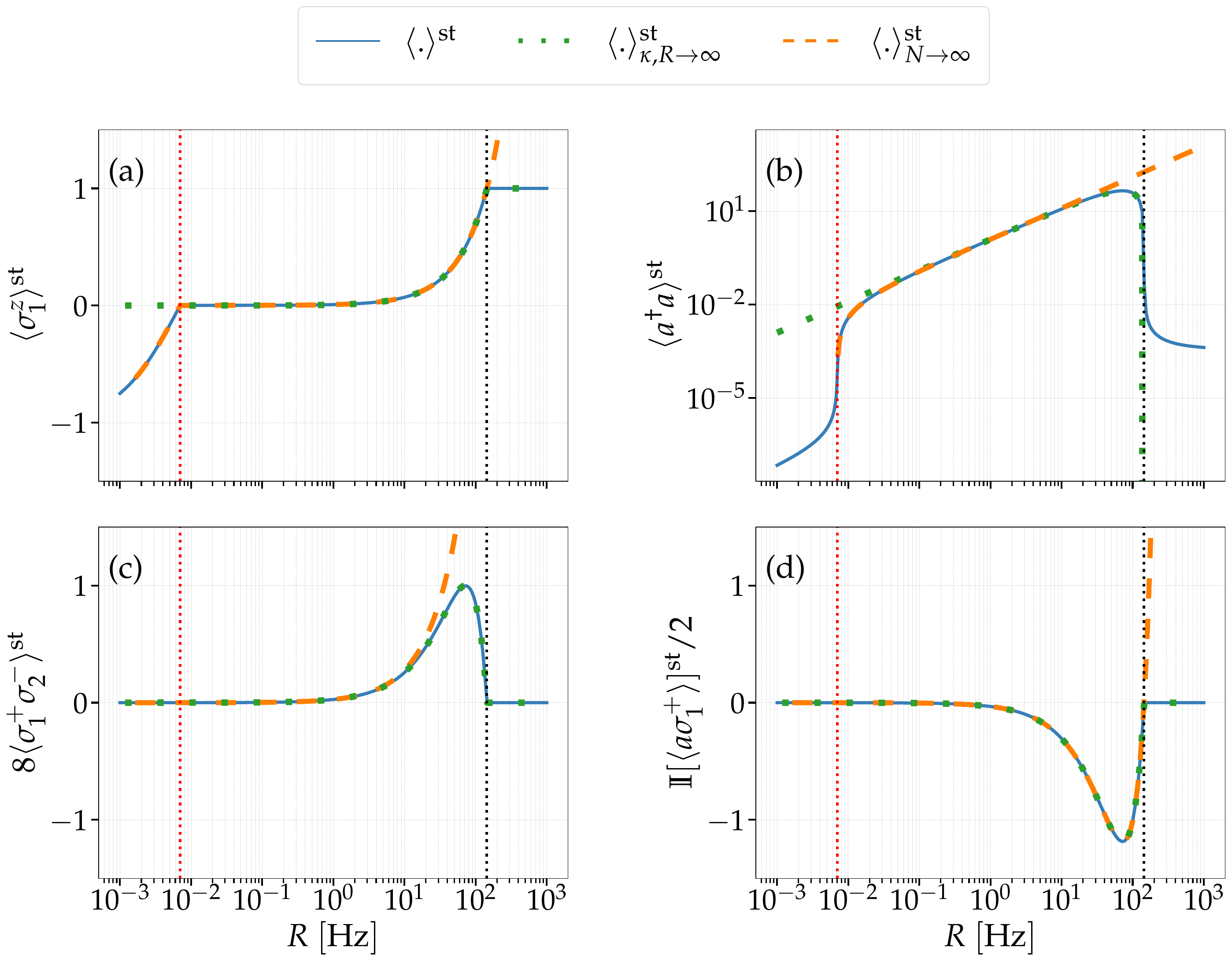}
	\caption{Steady-state solutions of Eqs.~\eqref{eqs:Steady_state} as a function of $R$ for $N=10^6$ atoms represented by solid lines. The dotted lines refer to the large $\kappa$ and $R$ limit described in Appendix B, while dashed lines correspond to the large $N$ limit reported in Appendix C. Red (resp. black) vertical dotted line: $R_\mathrm{min}$ (resp. $R_\mathrm{max}$). }
	\label{2x2obs}
\end{figure}

For a laser operating in the bad-cavity limit, the superradiant emission in the system is characterized by both a lower and upper threshold with respect to $R$ \cite{Meiser2009}. The dependence of the various variables with respect to $R$ is presented in Figure~\ref{2x2obs} for $N = 10^6$ atoms. The lower threshold corresponds to the point at which the repumping rate overcomes the atomic decay rate $R_{\min } \ge \gamma$, which is necessary to generate the population inversion. From Eqs.~\eqref{Steady state relations} it follows that $\braket{\sigma^z_1}^{\mathrm{st}} < 0$ for $R<\gamma$, as it can be observed in Figure~\ref{2x2obs}(a). For such values of $R$ where the population inversion is not reached, the intracavity photon number is observed in Figure~\ref{2x2obs}(b) to be very small. Above this threshold $\braket{\sigma^z_1}^{\mathrm{st}}$ crosses zero and $\braket{a^\dagger a}^\mathrm{st}$ shows a sharp increase, indicating the onset of the superradiant emission. The atom-atom correlations $\braket{\sigma^+_1\sigma^-_2}^{\mathrm{st}}$ in Figure~\ref{2x2obs}(c) and the imaginary part of the atom-field correlations $\braket{a\sigma^+_1}^{\mathrm{st}}$ (real part is equal to zero for $\Delta = 0$) in Figure~\ref{2x2obs}(d), are close to zero for such small values of $R$. 

As the value of $R$ increases, all the above variables increase in modulus. In particular, the photon number increases linearly with $R$, as found in Eqs.~\eqref{ada large N} in the large $N$ limit for $R>\gamma$. 
Based on the approximate Eqs.~\eqref{Steady state relations simple large N} in the large $\kappa$ and $R$ limit, we observe that the atom–atom correlations and the photon number both reach a maximum at $R= 2g^2N/\kappa$, with $\braket{\sigma_i^+ \sigma_j^-}^{\mathrm{st}} \approx 1/8$ and $\braket{a^{\dagger}a}^{\mathrm{st}} \approx g^2N^2/(2\kappa^2)$. This value of $R$ is then defined as the optimal rate, $R_{\mathrm{opt}} \equiv  2g^2N/\kappa$, where superradiance manifests the most. At $R_\mathrm{opt}$, it holds $\braket{\sigma^z_1}^{\mathrm{st}} \approx 1/2$, and the atom-field correlations reach a minimum, $\braket{a \sigma^+_1}^{\mathrm{st}} \approx -gN/(4\kappa)$. 
For values of $R$ above this optimal rate, $\braket{\sigma^z}^\mathrm{st}$ approaches 1 for $R =2 R_{\mathrm{opt}}= 4g^2N/\kappa$, and maintains this value for larger values of $R$, as shown in Eqs.~\eqref{Steady state relations small N} in the large $\kappa$ and $R$ limit. At $2 R_{\mathrm{opt}}$, the photon number, the atom-atom correlations, and the atom-field correlations are all close to zero. This value of $R$ is then the upper threshold of the superradiant emission and it is defined as, $R_{\mathrm{max}} \equiv 2 R_{\mathrm{opt}}=4g^2N/\kappa = NC\gamma$, where the cooperativity parameter $C$ is defined as $C = 4g^2/ (\kappa\gamma)$ \cite{Kazakov2022}. The term $C\gamma$ is the single-atom decay rate defined in the atomic master equation following the adiabatic elimination of the cavity mode in the bad-cavity limit \cite{intensity_Holland}. 
It follows that $NC\gamma$ characterizes the $N$-atom dynamics. The condition $R \ge R_{\max }$ then corresponds to the case when the repumping rate is equal or larger than this collective decay rate.

In Figure~\ref{2x2obs} we also compare the solutions of Eqs.~\eqref{eqs:Steady_state} with their approximate expressions for the large $\kappa$ and $R$ limit and the large $N$ limit. The large $\kappa$ and $R$ limit is accurate for the atom-atom and the atom-field correlations for all values of $R$. This is not the case for the atomic inversion and for the photon number. For the atomic inversion, it holds for $R\ge R_{\mathrm{min}}$, and for the photon number for $R^* \le R\le R_\mathrm{max}$, where $R^*$ is a certain value above $R_{\mathrm{min}}$. The large $N$ approximation is well satisfied for all quantities up to some value of $R$, depending on the value of $N$ we are considering. For the photon number at $R\le R_{\mathrm{min}}$, the solution becomes unstable for values close to $R_\mathrm{min}$, and for this reason the corresponding region is not shown in the figure.

\begin{figure}[b!]
	\centering
	\includegraphics[width=0.7\linewidth]{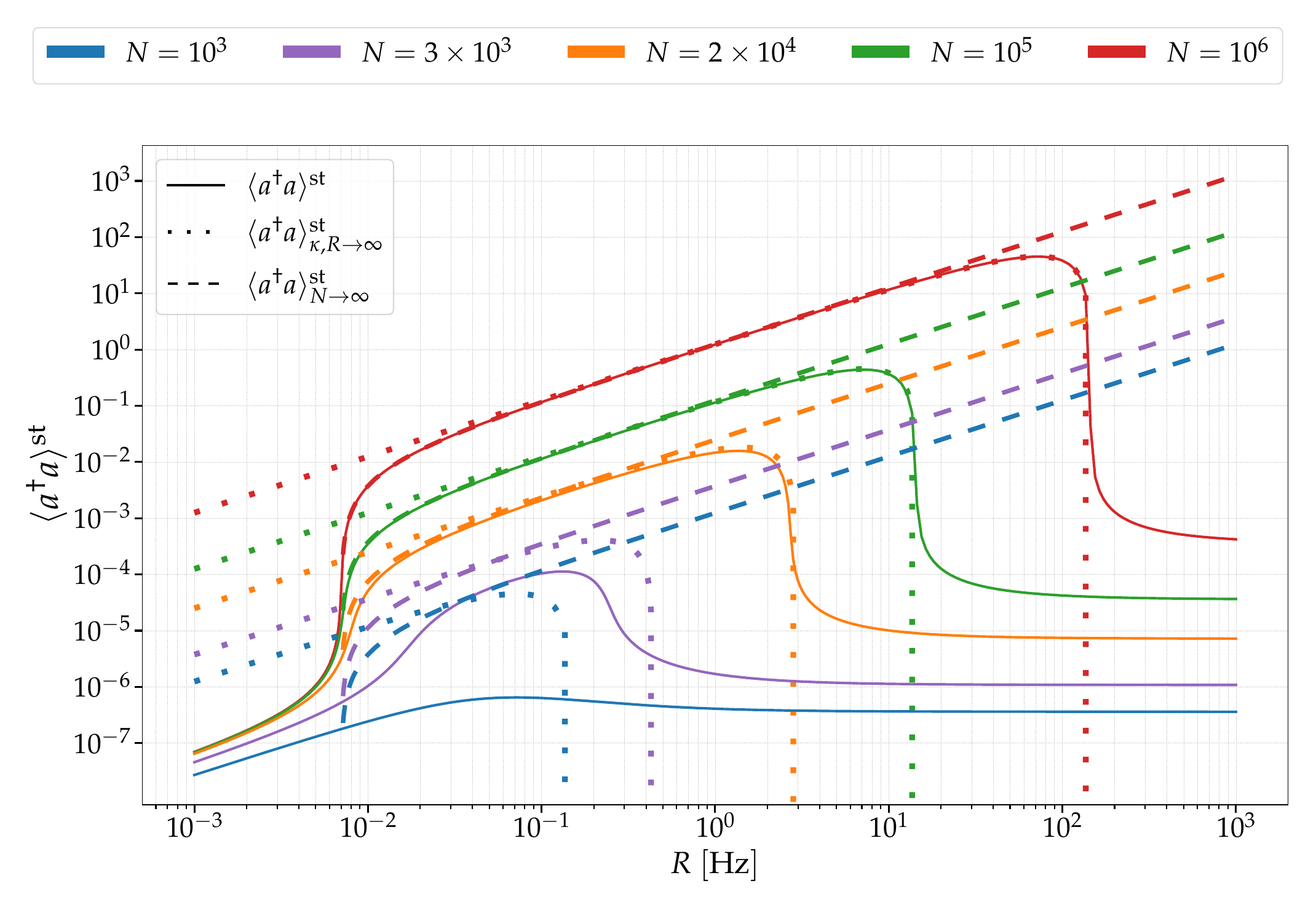}
	\caption{Number of intracavity photons at the steady-state $\braket{a^{\dagger}a}^{\mathrm{st}}$  as a function of the repumping rate $R$ for various values of $N$. The solid lines are the steady-state solutions of Eqs.~\eqref{eqs:Steady_state}, while the dotted lines refer to the large $\kappa$ and $R$ limit reported in~\autoref{app:simple} and the dashed lines correspond to the large $N$ limit reported in~\autoref{app:largeN}.}
	\label{fig:steady variables R}
\end{figure}

In order to observe the behavior of the photon number with the change in the atom number, we compare in Figure~\ref{fig:steady variables R} $\braket{a^{\dagger}a}^{\mathrm{st}}$ for several values of $N$, with its approximate expressions for the large $\kappa$ and $R$ limit, and the large $N$ limit. As observed above, the two limits can be well satisfied for certain values of $R$. This is the case for $N=10^5$ and $10^6$, while the agreement degrades more and more for smaller values of $N$.
We also notice that the maximum photon number when we move from $N=10^5$ to $10^6$, is increased by around two orders of magnitude, as expected from its quadratic scaling with $N$. 

%%%%%%%%%%%%%%%%%%%%%%%%%%%%%%%%%%%%%%%%%%%%%%%%%%%%%%%%%
%%%%%%%%%%%%%%%%%%%%%%%%%%%%%%%%%%%%%%%%%%%%%%%%%%%%%%%%%

\begin{figure}[b!]
	\centering
	\includegraphics[width=0.75\linewidth]{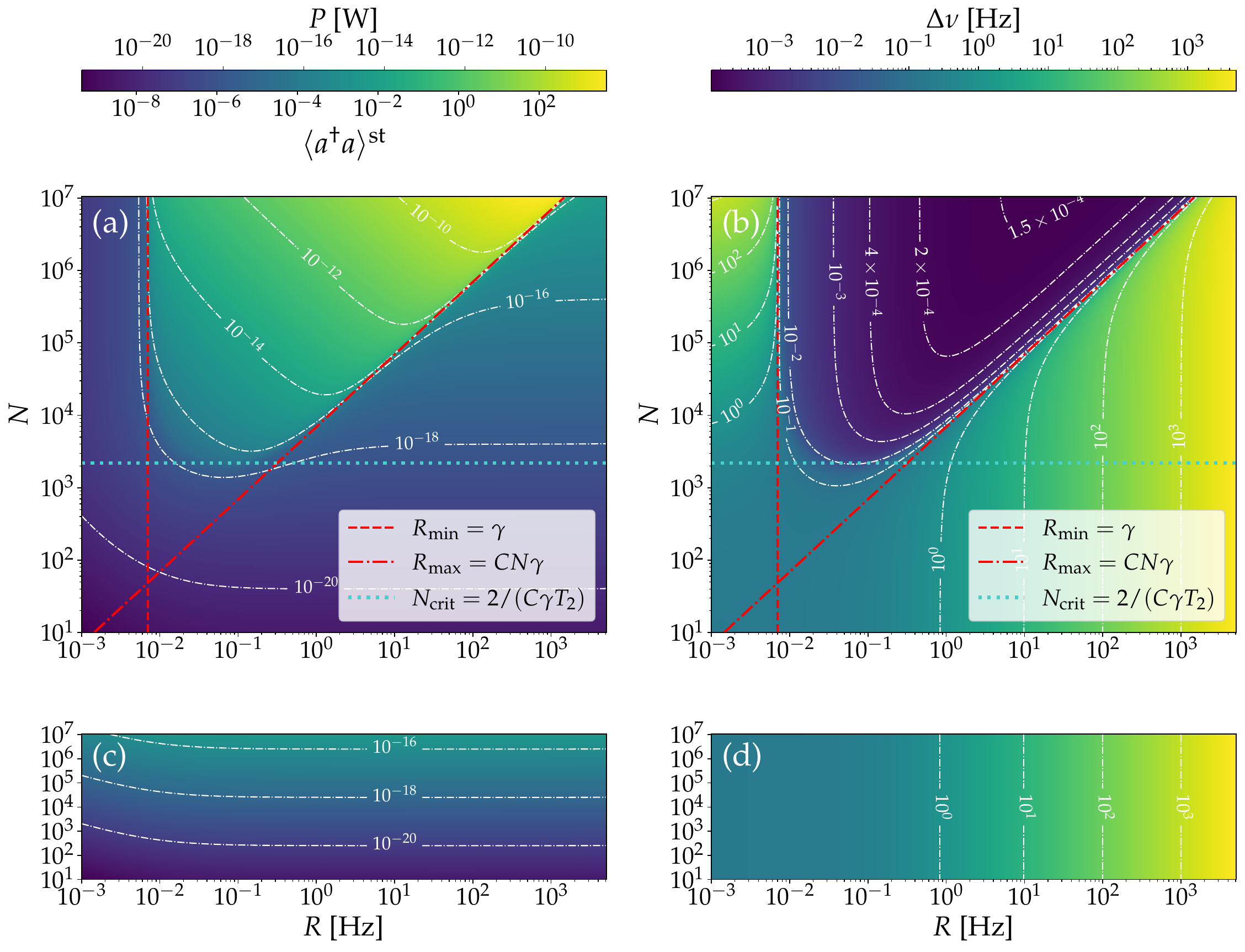}
	\caption{(a), (c) Intracavity photon number $\braket{a^\dagger a}^{\mathrm{st}}$ and output power $P$ and (b), (d) linewidth $\Delta\nu$ both as a function of atom number $N$ and repumping rate $R$. 
		(a)-(b) Collective dynamics with $N$ atoms coupled to a single cavity mode. Red dashed line: minimum repumping rate $R_{\mathrm{min}} = \gamma$ for superradiant emission onset; red dot-dashed line: maximum repumping rate $R_{\mathrm{max}} = N C \gamma$ above which superradiance is suppressed; light blue dotted line: critical atom number $N_\mathrm{crit} = 2/(C \gamma T_2)$ for the onset of collective behavior \cite{Meiser2009}. (c)-(d) Non-collective dynamics of $N$ single-atom systems each coupled to an individual cavity with identical parameters as in (a)-(b). Various contour lines (linking points with the same value) are depicted by white dot-dashed lines.}
	\label{fig:power_and_linewidth_1class}
\end{figure}

The study of the steady-state solutions of the atomic and cavity variables pointed out that around $R_{\mathrm{opt}}$ the system could be exploited as a superradiant laser \cite{Meiser2009}. In Figure~\ref{fig:power_and_linewidth_1class}, we analyze the steady-state laser characteristics for a large range of $R$ and $N$. Figure~\ref{fig:power_and_linewidth_1class}(a) shows the photon number and the output power, while Figure~\ref{fig:power_and_linewidth_1class}(b) displays the linewidth of the cavity field spectrum, all as functions of $R$ and $N$. To show the importance of the common cavity mode in the generation of atom-atom correlations, we compare the collective configuration with an alternative one consisting of $N$ independent single-atom systems, each coupled to its own cavity with parameters identical to that of the $N$-atom case. The intracavity photons from all single-atom systems are summed, and the corresponding total output power is given in Figure~\ref{fig:power_and_linewidth_1class}(c). In this configuration, the linewidth is equal for all single-atom systems and it corresponds to the linewidth of the total system, which is given in Figure~\ref{fig:power_and_linewidth_1class}(d). In the presence of strong collective effects in the $N$-atom system, significant differences between the two configurations are expected.

In the collective dynamics with $R<R_{\mathrm{min}}$, the small number of intracavity photons results in very low output power, and the linewidth of the laser is found to be broad. Above $R_{\mathrm{min}}$ the sharp increase in the photon number translates into a pronounced increase in output power, and the linewidth becomes narrow as the atom-atom correlations are enhanced. 
For $R = R_{\mathrm{opt}}$, both the photon number and the atom-atom correlations reach a maximum. 
This produces the maximum power $P_{\mathrm{max}} = \hbar \omega_c N^2C\gamma/8$, as reported in  ~ \cite{Meiser2009, Kazakov2022, Bohnet2012, koreaN^2}, and the linewidth is close to the minimum expected for superradiant lasers $\Delta \nu \approx C \gamma$, as also found in \cite{Meiser2009, Kazakov2022} and recovered in Eq.~\eqref{app:FWHM special case form large N}.
Above the upper threshold $R_{\mathrm{max}}= NC\gamma$, both the photon number and the atom-atom correlations decreases, and this reduces the power. For the same values of $R$, the linewidth is observed to be broad. The relevant range of repumping rates for a superradiant laser is then $R_{\mathrm{min}}<R<R_{\mathrm{max}}$. Within this range, and for large values of $N$, the linewidth is nearly independent of $N$ up to certain values of $R$, as shown by the contour lines in Figure~\ref{fig:power_and_linewidth_1class}(b), and  commented in \autoref{app:spectrum} based on Eq.~\eqref{app:ata:spectrum:N}. 
With a reasonable atom number for this experiment, $N = 10^6$, the laser output can reach $P_{\mathrm{max}} \approx 31~\mathrm{pW}$ with a corresponding linewidth of $\Delta\nu \approx 0.35~\mathrm{mHz}$.

For the case of the single-atom systems shown in Figure~\ref{fig:power_and_linewidth_1class}(c)-(d), the output power increases linearly with the number of atoms $N$. Above the threshold $R>R_{\mathrm{min}}$, this power shows negligible variation with an increase in $R$. The linewidth for this system is broad at small $R$ and it increases linearly with $R$. This behavior is in contrast to that observed in the collective dynamics for large $N$, where the power scales with $N^2$ and the linewidth becomes very narrow. This clearly shows the importance of collective coupling, enabled by the common cavity mode, in order to have a superradiant emission process that enhances the laser performance. 

With respect to the atom number impact on the dynamics, it was found in \cite{Meiser2009} that in the limit of $T_2 \gamma \ll 1$, there is a critical number of atoms $N_{\text {crit }}=2 /\left(C \gamma T_2\right) \approx 2.2 \times 10^3 $ required  to reach the collective dynamics of the superradiant laser. This critical number is marked in Figure~\ref{fig:power_and_linewidth_1class}(a)-(b) by a light blue line, and for $N<N_\mathrm{crit}$ of this figure, the system shows a behavior similar to the single-atom systems in Figure~\ref{fig:power_and_linewidth_1class}(c)-(d).
In the parameter space where the collective dynamics is reached, the superradiant laser linewidth is demonstrated to be remarkably insensitive to cavity dephasing $\xi$ over a broad range of values in the theoretical study of \cite{Bychek_2021}. To confirm this for our system, we introduced a broad range of $\xi$ values in the model and observed no significant changes for values up to approximately $\xi = 10^4 \mathrm{~Hz}$.

\section{Inhomogeneous broadening and variable coupling strength}\label{Xclass}

\begin{figure}[h!]
	\centering
	\includegraphics[width=0.7\linewidth]{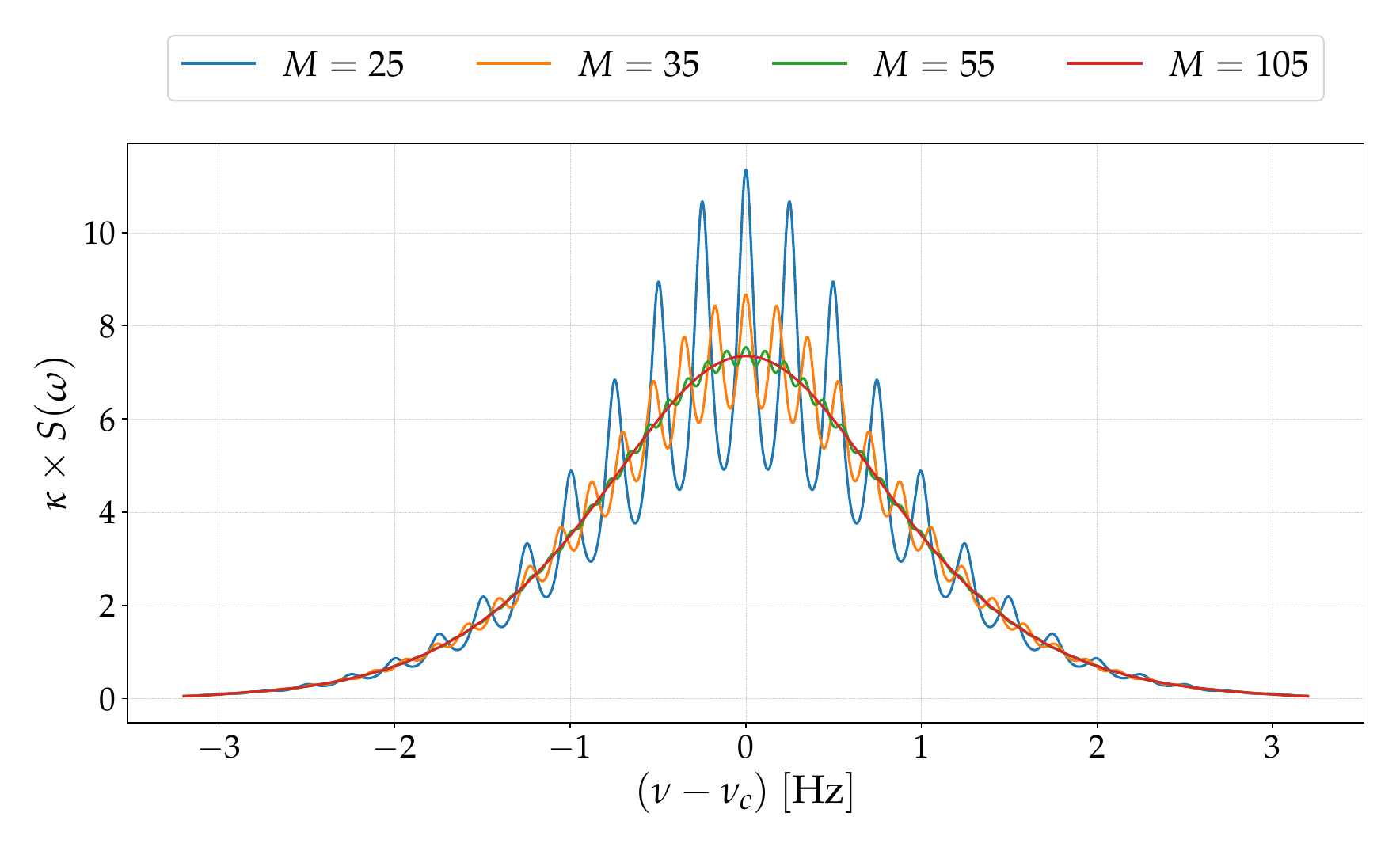}
	\caption{Cavity spectra for $N = 2 \times 10^4$ atoms, $\sigma = 2\pi \times1~\mathrm{Hz}$ standard deviation, and $R = 2\pi\times0.01 ~\mathrm{Hz}$, for various values of the number of classes $M$. }
	\label{fig:clusters}
\end{figure}

\begin{figure}[t!]
	\centering
	\includegraphics[width=0.7\linewidth]{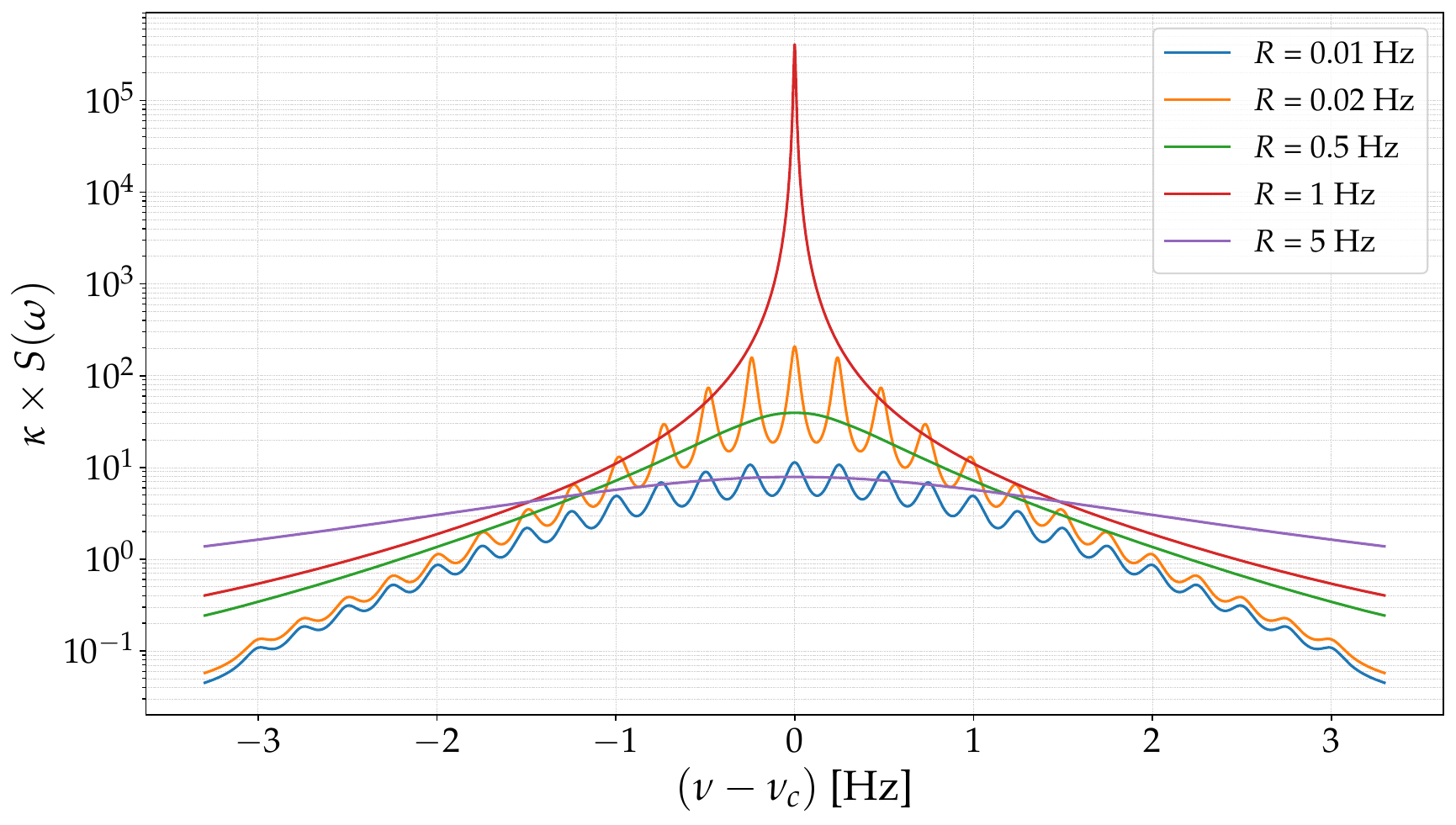}
	\caption{Cavity spectra for $M=25$ classes, $N=2\times 10^4$ atoms, and $\sigma =2\pi\times 1 ~\mathrm{Hz}$, for several values of $R$. }
	\label{fig:sync_M_classes}
\end{figure}

\begin{figure}[h!]
	\centering
	\includegraphics[width=0.7\linewidth]{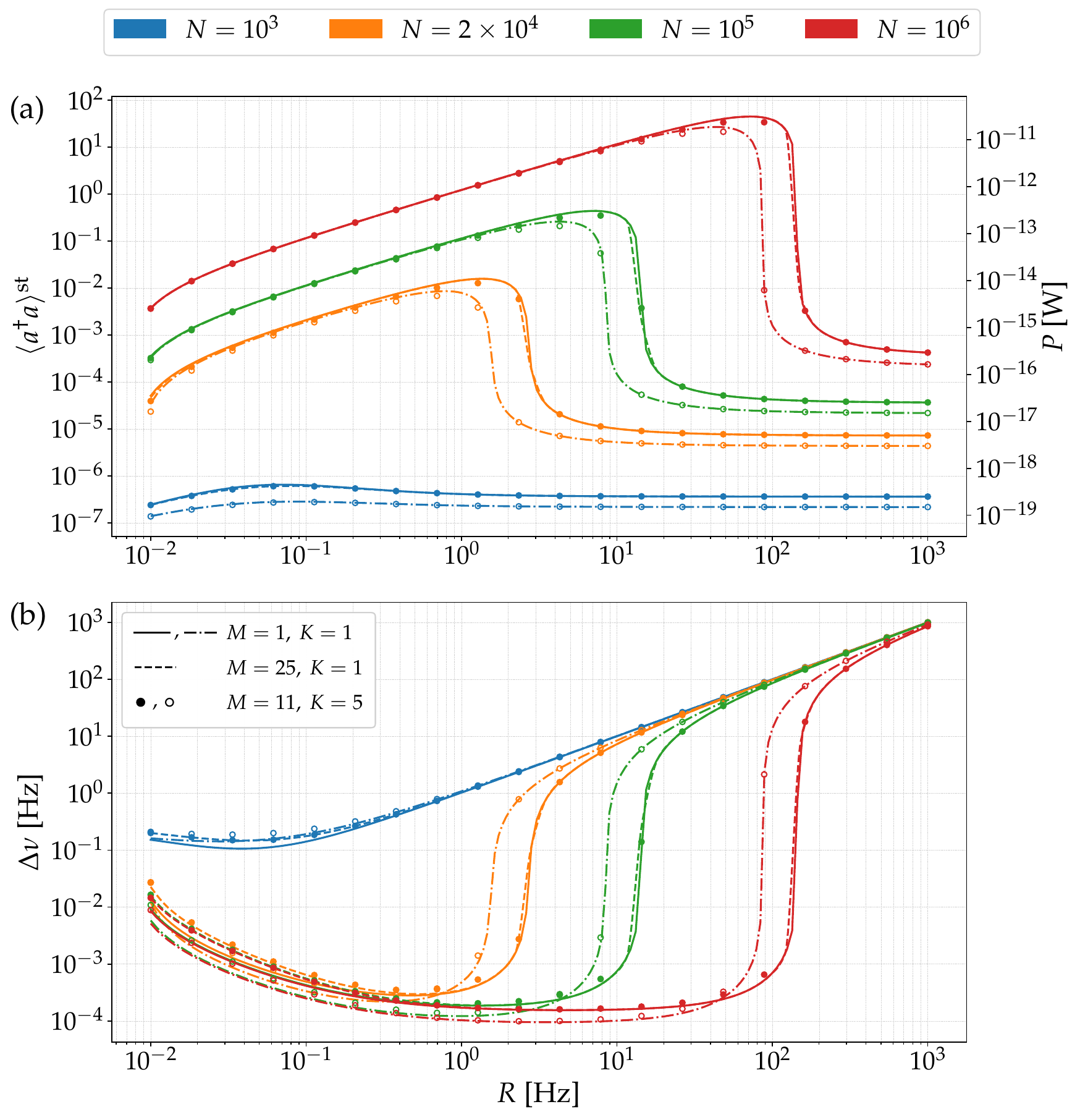}
	\caption{For several total atom numbers we have: (a) number of intracavity photons at the steady-state $\braket{a^\dagger a}^\mathrm{st}$ and output power $P$, and (b) linewidth of the cavity spectrum $\Delta\nu$ as a function of $R$. Solid and dot-dashed lines correspond to the reference of one-class case which corresponds to $\Delta\nu_{\mathrm{m}} = 0$, and $g = 2\pi\times 3.8~\mathrm{Hz}$ for the solid lines and $g = 2\pi\times 2.9~\mathrm{Hz}$ for the dot-dashed lines. Dashed lines correspond to the $M = 25$ case with $\Delta\nu_{\mathrm{m}} = 2\pi\times0.1 ~\mathrm{Hz}$, and $g = 2\pi\times 3.8~\mathrm{Hz}$. Circles correspond to the $M = 11$ and $K=5$ case with $\Delta\nu_{\mathrm{m}} = 2\pi\times0.1 ~\mathrm{Hz}$ and varying  $g(x)$: full circles correspond to $g_0 =2\pi\times 4.9 ~\mathrm{Hz}$, and empty circles correspond to $g_0 = 2\pi\times 3.8~\mathrm{Hz}$.}
	\label{fig:M_classes} 
\end{figure}

From the simple case of identical atoms equally coupled to the cavity considered in \autoref{1class}, we now move towards a more realistic description of the atomic ensemble used in superradiant lasers. In experimental implementations, atoms typically experience variations in their resonance frequencies due to inhomogeneous broadening, which can arise from magnetic field gradients or local environmental fluctuations. This broadening leads to a distribution of resonance frequencies across the ensemble of atoms. In addition to a frequency distribution, it is also important to account for variations in the atom-cavity coupling strengths. Such variations arise from the spatial profile of the cavity mode. Atoms located at different positions experience variations in the field amplitude and therefore couple to the cavity with different strengths.
To account for both variations, we first consider a frequency distribution across the atoms in Figures~\ref{fig:clusters} and~\ref{fig:sync_M_classes}, and then extend the description to include variations in the coupling strength in Figure~\ref{fig:M_classes}. For this we follow the approach described in \autoref{subsec:2ndcumulant}.

An atomic ensemble composed of $M$ discrete frequency classes, each characterized by a detuning $\Delta_{m}$, $m=1,\ldots$,$M$, spaced uniformly over a given frequency interval, is considered in Figures~\ref{fig:clusters} and~\ref{fig:sync_M_classes}. The number of atoms in each class is sampled from a Gaussian distribution in order to capture the statistical nature of the inhomogeneous broadening in the system \cite{Bychek_2021}. This distribution is characterized by a FWHM $\Delta \nu_{\mathrm{m}} = 2\sqrt{2\ln(2)} \sigma$, where $\sigma$ is the corresponding standard deviation. In order to capture the significant part of the Gaussian profile, the frequency classes are chosen within the interval $[-3\sigma,~3\sigma]$. 

The spectrum of the cavity field is shown in Figure~\ref{fig:clusters} for $N = 2 \times 10^4$ atoms, with $\sigma = 1 ~\mathrm{Hz}$, and $R=0.01 ~\mathrm{Hz}$ which is just above the lower threshold. For a small number of classes, $M = 25$, the spectrum displays a set of distinct spectral lines corresponding to the discrete frequency classes. When $M$ increases, the distinct lines start to average out in the spectrum. The spectrum approaches a continuous profile of a single broad emission line centered at the cavity frequency $\nu_c = \omega_c/2\pi$ for $M = 105$, which reflects the behavior of the actual physical system. The linewidth of this spectrum for the given parameters is $\Delta\nu \approx 1.92 ~\mathrm{Hz}$ that is very close to the FWHM of the distribution $\Delta\nu_m \approx 2.4 ~\mathrm{Hz}$. To ensure reasonable computational times, we therefore limit the following analysis to a moderate value of $M = 25 $. While this choice does not fully reproduce the physical spectrum, it allows us to qualitatively capture the overall spectral behavior expected in the limit of a very large number of frequency classes.

With the number of frequency classes set to $M = 25$, we now examine how variations in the repumping rate can affect the cavity spectrum. In the one-class case, we observed that increasing $R$ enhances atom-atom correlations leading to a narrow spectral line around $R_{\mathrm{opt}}$. The same behavior is observed here in Figure~\ref{fig:sync_M_classes} for $M=25$ frequency classes. Starting from $R= 0.01 ~\mathrm{Hz}$, the spectrum shows several spectral lines corresponding to the various $\Delta_m$. As the repumping rate is increased to $R=0.5 ~\mathrm{Hz}$, the spectrum has a single peak centered at the cavity frequency, and the spectral linewidth reduces to $\Delta\nu \approx 1.0 ~\mathrm{Hz}$, which is smaller than the FWHM of the distribution $\Delta\nu_m \approx 2.4 ~\mathrm{Hz}$. At $R= 1~\mathrm{Hz}$, the spectral line becomes more pronounced and it is characterized by a narrow linewidth $\Delta\nu \approx 0.012 ~\mathrm{Hz}$. This kind of behavior indicates the occurrence of synchronization among atoms having different detuning $\Delta_m$ \cite{Bychek_2021,Minghui_NA=NB}. When the repumping rate is further increased to $R = 5 ~\mathrm{Hz}$, the spectrum becomes broader than $\Delta\nu_m$, having a linewidth $\Delta\nu \approx 3.2 ~\mathrm{Hz}$.
This shows that the spectral behavior for a system with $M$ frequency classes follows qualitatively that of a system with one-class. For a one-class system with the same parameters as in Figure~\ref{fig:sync_M_classes}, the optimal repumping rate is $R_{\mathrm{opt}} \approx 1.4 ~\mathrm{Hz}$, consistent with our observation in Figure~\ref{fig:sync_M_classes}.
For the above study, we set $\sigma = 1~\mathrm{Hz}$ which corresponds to $\Delta\nu_m \approx 2.4 ~\mathrm{Hz}$. This was done in order to clearly show distinct spectral lines for small values of $M$. 

We now move to a more experimentally reasonable situation and consider $\Delta\nu_m = 0.1 ~\mathrm{Hz}$, to examine how this affects the spectral properties of the system. In addition, we now  account for variations in the atom–cavity coupling strength, deriving from the position dependence of the atom-cavity interaction. This framework allows us to we compare the steady-state laser characteristics for three different descriptions of the system. The first one is a reference case, which corresponds to the one-class case. The second corresponds to a system with all couplings equal to that of the one class case, this time with a frequency broadening treated with $M = 25$ classes. The last description corresponds to a system with frequency broadening and variations in the atom-cavity coupling strengths treated, respectively, with $M=11$ and $K=5$ classes. This comparison is done in Figure~\ref{fig:M_classes}, where the output power is shown in  Figure~\ref{fig:M_classes}(a) and the corresponding linewidth in  Figure~\ref{fig:M_classes}(b), each as a function of $R$ for several total number of atoms $N$. 

The variation in the coupling strengths is described by $g(x) = g_0\mathrm{cos}(kx)$, where $x$ is the atomic position within the cavity, $g_0$ is the maximum coupling strength, and $k = 2\pi/\lambda$ is the cavity mode wave number \cite{Bychek_2021}. 
The atoms are divided equally into $K$ strength classes, where each class corresponds to a value of $g(x)$ obtained from a uniform distribution of positions in the interval $x \in$ $[0, ..., \lambda/4)$. Since the dynamics of the system only depends on the magnitude of the coupling strength, we limit the analysis to positive values of $g(x)$. As observed in \cite{Bychek_2021}, systems with variable coupling strengths can reproduce the results of the one-class system by using an effective value equal for all the atoms, which is given by $g_{\mathrm{eff}} = \sqrt{\sum_{j =1}^K g(x_j)^2/K}$. 
The value $g = 3.8~\mathrm{Hz}$ considered in \autoref{1class} actually corresponds to the maximum achievable in our setup. If we take $g_0 = g = 3.8~\mathrm{Hz}$, however, the effective strength no longer matches the one-class value, $g_{\mathrm{eff}}\approx 2.9~\mathrm{Hz}$. To be consistent with the one-class case at $g = 3.8~\mathrm{Hz}$ (solid line in the figure), we set the maximum coupling to $g_0 \approx 4.9~\mathrm{Hz}$, giving $g_{\mathrm{eff}} = 3.8~\mathrm{Hz}$ (full circles). We also present the experimentally relevant case of frequency broadening with variable couplings at $g_0 = 3.8~\mathrm{Hz}$ (empty circles), together with the corresponding one-class case $g = g_{\mathrm{eff}} \approx 2.9~\mathrm{Hz}$ (dot-dashed line). The case with frequency broadening and uniform couplings is considered only for $g = 3.8~\mathrm{Hz}$ (dashed line).

Starting in Figure~\ref{fig:M_classes} with the frequency broadened system treated with $M=25$ classes, we find that it reproduces the output power of the reference one-class system. The impact of frequency broadening becomes apparent in the linewidth, where at low repumping rates the system with $M=25$ shows a slightly larger value compared to the one-class case. As the repumping rate increases and approaches the FWHM of the frequency distribution $\Delta\nu_\mathrm{m}$, the linewidth of the broadened system approaches that of the one-class reference. For a total number of atoms above $N_{\mathrm{crit}}$ defined in the one-class case, the laser reaches a sub-millihertz linewidth. 
Concerning the systems with both frequency broadening and variation in coupling strengths, we can observe that it also follows the power of the corresponding one-class system. For the system with $g_\mathrm{eff} = 3.8~\mathrm{Hz}$, comparison with the case including only frequency broadening shows that variations in the coupling strength have almost no additional effect on the linewidth. 
Comparing the cases with effective couplings of approximately $3.8~\mathrm{Hz}$ and $2.9~\mathrm{Hz}$, we find that the upper threshold of the superradiant emission, the optimal repumping rate for maximal output power, and the maximum power are all reduced. This behavior is expected since these quantities scale with the square of the effective coupling. In particular, the shift in the systems response with respect to $R$ for $g_\mathrm{eff} \approx 2.9~\mathrm{Hz}$ comes from the smaller upper threshold in this case. The minimum linewidth is also narrower, since it scales with the cooperativity $C$, which itself scales with the square of the effective coupling.

With fixed parameters and a scan over $R$, the optimal operating point for the laser in terms of output power occurs at $R_\mathrm{opt}$. At this point, the linewidth reaches a sub-millihertz value that is close to the minimum linewidth $\Delta\nu = C\gamma$ defined with the one-class system. 
Figure~\ref{fig:M_classes} shows that in this parameter range, the laser performance is robust to both inhomogeneous broadening and coupling strength variations, provided that the effective coupling matches that of the reference case.

\section{Two-site setup}\label{sec:2sites}

We now examine the impact of having two atomic ensembles at separate sites on the superradiant emission. In the FEMTO-ST experiment described in \autoref{sec:femto}, it is planned to have two sites $A$ and $B$ distant by approximately 3~mm. Each site is populated with atoms sequentially, in order to compensate for the decay of atoms from the cavity and ensure a continuous superradiant emission. As a consequence of magnetic field gradients or other local environmental fluctuations, the two ensembles are subject to different frequency shifts. In order to include this in the description of the system, we study the cavity spectrum in the case when all the atoms in $A$ have the same detuning $\Delta_A$ and the atoms in $B$ have the same detuning $\Delta_B$. The study is limited to the case $\Delta_A = - \Delta_B = \Delta$, which could be realized tuning properly the cavity frequency. Based on the conclusion from \autoref{Xclass} concerning the variation of the coupling strengths, we consider in this section that the atoms are all equally coupled to the cavity, which means $K=1$. In \autoref{section:NA=NB}, we consider the case where we have a balanced atom number $N_A=N_B$ and study the corresponding cavity spectrum. Then we move to the imbalanced case $N_A\neq N_B$ in \autoref{section:NAdiffNB}, which is the configuration expected experimentally as a result of the sequential transport procedure.

\subsection{Balanced atom number: \texorpdfstring{$N_A = N_B$}{NA=NB}} 
\label{section:NA=NB}

\begin{figure}[h!]
	\centering
	
	\begin{subfigure}[t]{0.495\linewidth}
		\centering
		\includegraphics[width=\linewidth]{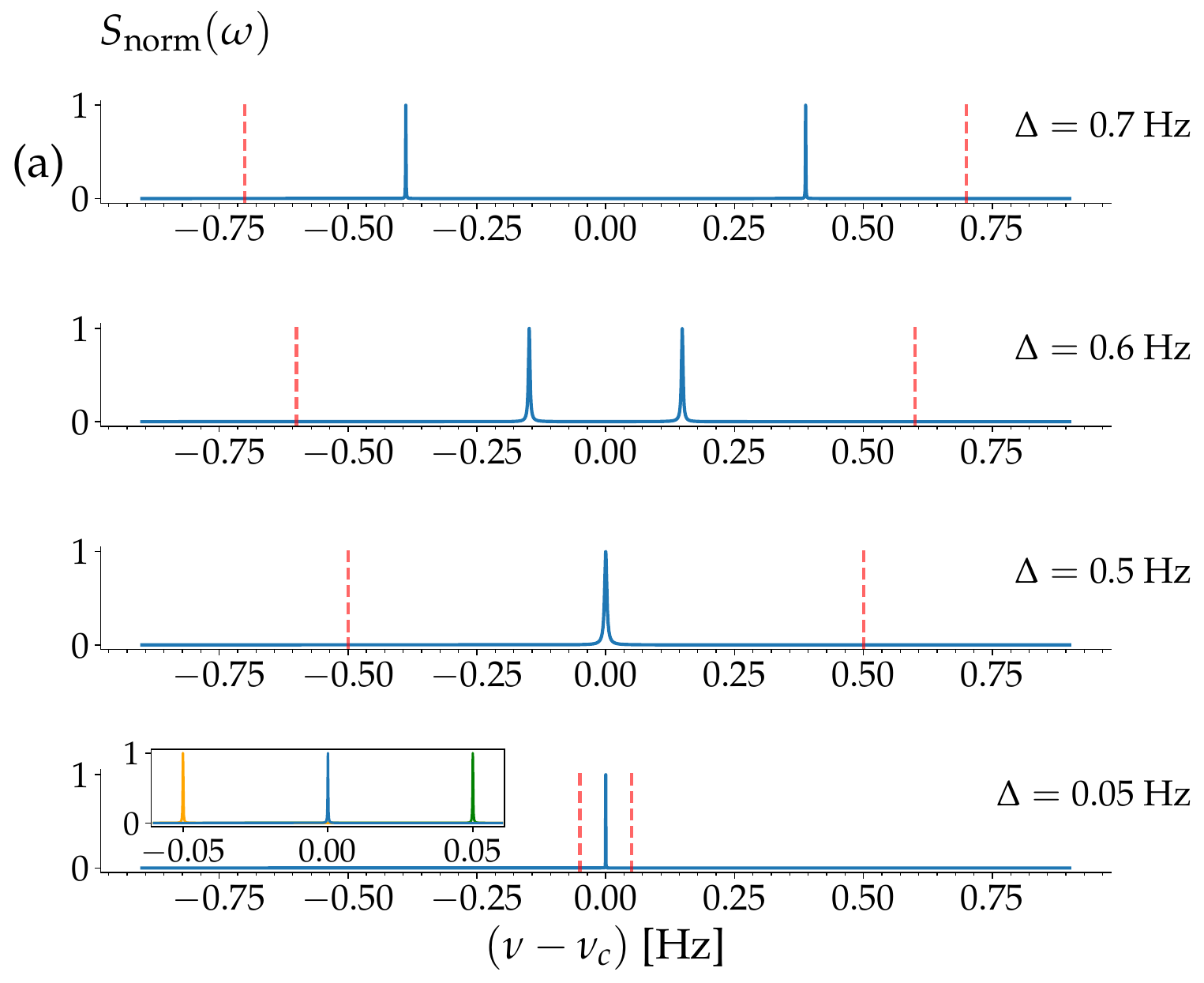}
	\end{subfigure}
	\hfill
	\begin{subfigure}[t]{0.495\linewidth}
		\centering
		\includegraphics[width=\linewidth]{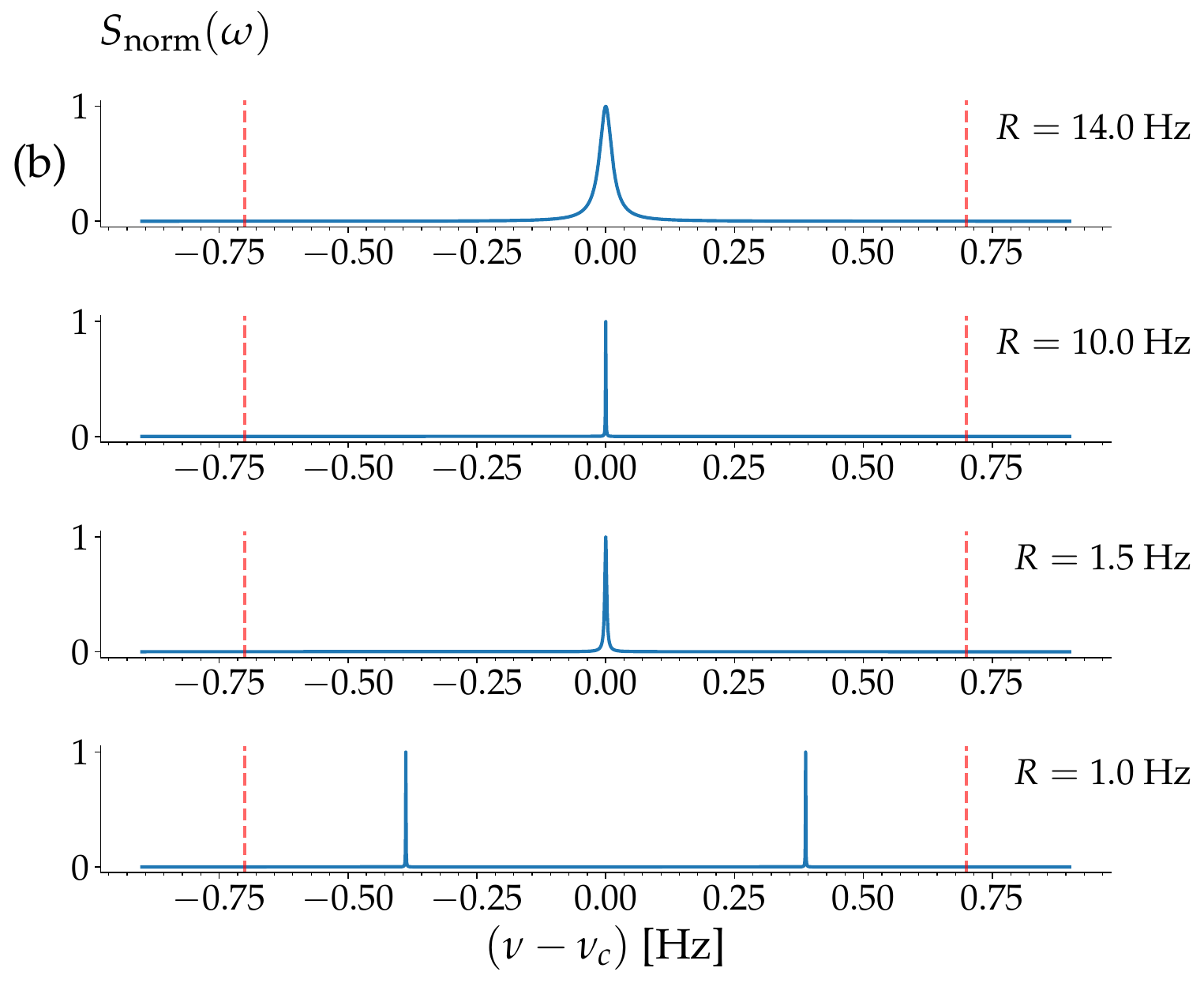}
	\end{subfigure}
	
	\caption{Normalized cavity field spectrum in the case of two atomic ensembles at site $A$ and $B$ with $N_A = N_B = 5\times10^4$ atoms, (a) for $R = 2\pi\times 1~\mathrm{Hz}$ and several values $\Delta$ from $2\pi\times0.05~\mathrm{Hz}$ to $2\pi\times0.7~\mathrm{Hz}$. Lower panel inset: cavity field spectrum in the presence of atoms only at site $A$ ($B$) given by the yellow (green) curve. (b) For $\Delta = 2\pi\times0.7~\mathrm{Hz}$ and several values of $R$ from $2\pi\times1~\mathrm{Hz}$ to $2\pi\times14~\mathrm{Hz}$. The values $\Delta_A=-\Delta_B =\Delta$ is indicated by the dashed red lines.}
	\label{fig:subplots2classes}
\end{figure}

We first consider the case of a balanced atom number $N_A=N_B$. A theoretical study of this configuration was performed in \cite{Minghui_NA=NB}, where the fast cavity decay relative to other processes in a superradiant laser was used to adiabatically eliminate the cavity field and reduce the description to atomic operators. In our study, we work with the complete system of atoms and the cavity described with Eqs.~\eqref{eqs:Steady_state} treated with $M=2$ frequency classes.
To better observe the different regimes, we first work with a low atom number $N_A = N_B = 5\times10^4$. The corresponding spectra are indicated in Figure~\ref{fig:subplots2classes} as a function of $\Delta$ and $R$.  
Figure~\ref{fig:subplots2classes}(a) shows the spectrum for $R = 1~\mathrm{Hz}$ and $\Delta$ varying from $0.05~\mathrm{Hz}$ to $0.7~\mathrm{Hz}$. At the small value $\Delta = 0.05~\mathrm{Hz}$, the spectrum consists of a single spectral line characterized by a narrow linewidth. In the inset we provide the cavity field spectrum when only atoms at site $A$ (yellow curve) or site $B$ (green curve) are in the cavity. For each case, it corresponds to a narrow spectral line centered at the frequency of the atoms. This shows that the cavity field spectrum observed when both sites are populated, is the outcome of the synchronization of atoms in $A$ with atoms in $B$. Since the detunings are symmetric with the cavity frequency and the atom number is balanced between the two sites, the central frequency of the spectrum is at zero.  
With an increase in $\Delta$, the spectrum first broadens, as observed at $\Delta = 0.5 ~\mathrm{Hz}$, and then two narrow spectral lines are produced, as observed at $\Delta = 0.6 ~\mathrm{Hz}$. For a larger value $ \Delta = 0.7 ~\mathrm{Hz}$, the spectral lines move further apart and approach $\Delta$ and $-\Delta$. 
In Figure~\ref{fig:subplots2classes}(b) we have the spectrum for $\Delta = 0.7 ~\mathrm{Hz}$ and $R$ varying from $1~\mathrm{Hz}$ to $14~\mathrm{Hz}$. With this increase in $R$, the atoms synchronize and the two spectral lines reduce to a single spectral line as observed for $R = 1.5 ~\mathrm{Hz}$, and the corresponding linewidth becomes smaller as $R$ increases further to $10~\mathrm{Hz}$. For a larger value $R = 14 ~\mathrm{Hz}$, the spectrum becomes broad. Synchronization here is important in order to produce a single spectral line with a narrow linewidth. A qualitative remark from this observation is that, in order to have this spectral characteristics, here the repumping rate has to be comparable to $|\Delta_A-\Delta_B| = 2\Delta$.  

%%%%%%%%%%%%%%%%%%%%%%%%%%%%%

We now study the performance of this laser composed of $M=2$ frequency classes in terms of its output power and linewidth. 
\begin{figure}[h!]
	\centering
	\includegraphics[width=0.7\linewidth]{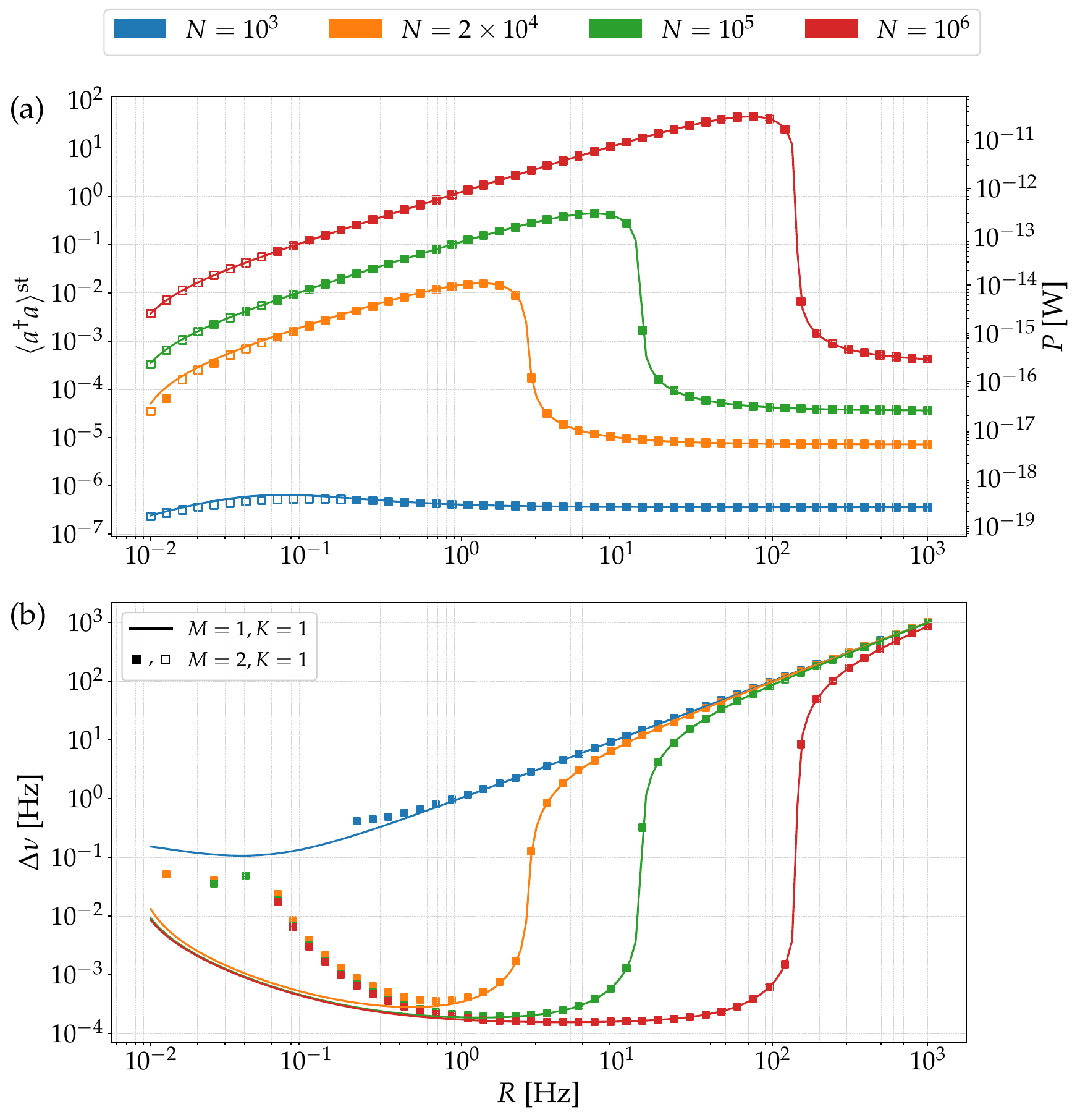}
	\caption{For several total atom numbers $N=N_A+N_B$ and  $N_A=N_B$ we have: (a) number of intracavity photons at the steady state $\braket{a^\dagger a}^\mathrm{st}$ and output power $P$, and (b) linewidth of the cavity field spectrum $\Delta\nu$ as a function $R$. Solid lines correspond to the reference of one-class case which corresponds to $\Delta\nu_{\mathrm{m}} = 0$ and $g = 2\pi\times 3.8~\mathrm{Hz}$. Squares correspond to the $M=2$ classes case with $\Delta = 2\pi\times0.1 ~\mathrm{Hz}$: full squares correspond to a spectrum with a single spectral line, and empty squares to that with more than a single spectral line.}
	\label{fig:2classes_P_and_nu}
\end{figure}
Figure~\ref{fig:2classes_P_and_nu} shows both quantities as a function of the repumping rate $R$ for $\Delta = 0.1~\mathrm{Hz}$, for several total atom numbers. The one-class case, corresponding to $\Delta=0$, is also given in the figure as reference case. The output power stays very close to that of the one-class case, as can be observed with the overlap of the squares that represent the $M=2$ classes with the solid lines of the one-class case. It is important to note that empty squares which appear for small values of $R$, correspond to the case where we observe two spectral lines in the system, and consequently the measurement of the linewidth is not recorded. The alternation in the occurrence of synchronization when $R$ increases is not a numerical artifact since the spectrum clearly shows either a single spectral line or two distinct ones. The figure also shows that this behavior depends on the number of atoms. Since the range of repumping rates at which this happens is not the one of interest here, we do not further analyze this phenomenon.

The linewidth for small values of $R$ is much larger than that of the one-class case. As the value of $R$ increases and approaches $2\Delta = 0.2~\mathrm{Hz}$, the linewidth gradually converges to that of the one-class case. This is a behavior similar to the one observed in Figure~\ref{fig:M_classes}, where the FWHM of the distribution $\Delta\nu_m$ of the $M$ classes was equal to $0.1 ~\mathrm{Hz}$. 
In the metrological superradiant laser, the values of $\Delta$ are expected to be at most in the $\mathrm{Hz}$ range, while the repumping rates will be tuned close to the optimal value $R_\mathrm{opt}$. In this parameter space, and for $N_A = N_B$, the system behaves very similarly to the single-class case.

%%%%%%%%%%%%%%%%%%%%%%%%%
\subsection{Imbalanced atom number: \texorpdfstring{$N_A \neq N_B$}{NA != NB}} 
\label{section:NAdiffNB}

\begin{figure}[h!]
	\centering
	\includegraphics[width=0.65\linewidth]{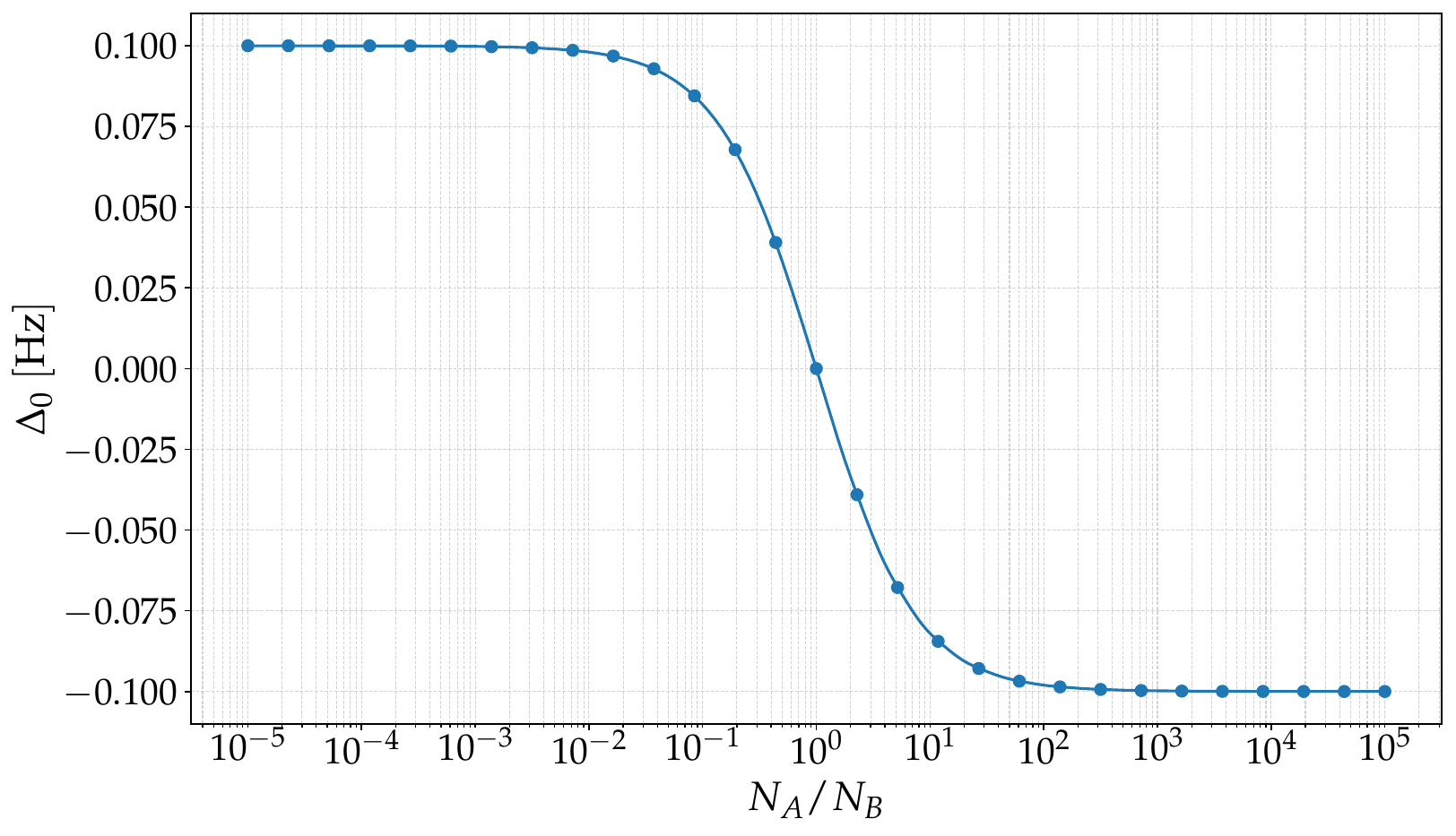}
	\caption{Shift of the central frequency of the cavity field spectrum with respect to the cavity frequency, $\Delta_0$, as a function of the atom number ratio $N_A/N_B$ for $R =2\pi\times 1 ~\mathrm{Hz}$ in the case of  $\Delta = 2\pi\times0.1 ~\mathrm{Hz}$ represented by full circles, compared to the weighted and shifted central frequency $\Delta_c= (N_A \Delta_A + N_B \Delta_B)/(N_A + N_B) $ represented by the solid line.}
	\label{fig:ratio_Na_Nb}
\end{figure}

We now turn to the experimentally relevant situation where $N_A$ and $N_B$ change in time due to both atom losses from the cavity and the sequential transport procedure, so that at a given moment $N_A \neq N_B$. We study the influence of the value $N_A/N_B$ on the central frequency of the spectrum and its output power by considering steady configurations for different values of this ratio. This approach is justified if the typical time for the atom number decay is much larger than the relaxation time of the system, which is expected to be the case experimentally.
The study is performed in a parameter space where the synchronization among atoms in the two sites happens, and this is the case for $\Delta = 0.1 ~\mathrm{Hz}$ at $R = 1~\mathrm{Hz}$.
With such parameters, the shift of the central frequency of the cavity field spectrum with respect to the cavity frequency, $\Delta_0$, is shown as a function of $N_A/N_B$ in Figure~\ref{fig:ratio_Na_Nb}. For comparison, we also show the weighted and shifted central frequency of the spectrum $\Delta_c = (N_A \Delta_A + N_B \Delta_B)/(N_A + N_B) $, with $\Delta_A$ and $\Delta_B$ the shifted frequency of atoms in sites $A$ and $B$, respectively, and find a very good agreement. In the parameter space we are working, this variation in the central frequency is independent of the total number of atoms. 
We have also computed the output power and the linewidth for this system and found that similar to what we have observed in \autoref{Xclass}, the power follows that of the one-class case with $N=N_A+N_B$, and that for a total number of atoms above the critical number defined with the one-class system, the linewidth can reach the mHz level in the optimal range of $R$ in the system.

For the sequential transport scheme of the superradiant laser developed at FEMTO-ST, \protect  Figure~\ref{fig:ratio_Na_Nb} shows that if the atoms in the two sites $A$ and $B$ do not share the same frequency, the central frequency of the laser shifts according to the relative population of the sites. 
This analysis can be used to qualitatively predict the behavior of the system in time, provided certain conditions are satisfied. If the typical timescale of atom loss and injection is much slower than the time it takes the system to reach a steady state, we can then use a coarse-graining approach: the total time is divided into intervals such that during each of them the system reaches so rapidly the steady state that we just consider it being in the steady state during all the given interval. In this way, it is meaningful to compute the cavity field spectrum at each interval and associate the various spectra with the real time dynamics of the laser. So from Figure~\ref{fig:ratio_Na_Nb}, we therefore expect that as atoms are sequentially loaded into or decay from the cavity, the central frequency of the laser oscillates from the detuning of one site to the other.

In the parameter space where synchronization occurs, the laser can reach a narrow linewidth as the atom number ratio varies, provided that $N_A+N_B$ is greater than the critical number of atoms $N_\mathrm{crit}$ defined in the one-class case. This demonstrates the importance of controlling the relative frequency of the two sites at a level consistent with the targeted frequency stability of the superradiant laser. From an experimental perspective, this sensitivity can be used as a diagnostic tool to measure the frequency difference and actively reduce it. If this difference cannot be reduced, an accurate knowledge of it allows for a compensation of this effect on the output signal.

%%%%%%%%%%%%%%%%%%%%%%%%%%%%%%%%%%%%%%%%%%%
\section{Conclusions and Perspectives}

We have conducted a theoretical study for the spectral properties of a superradiant laser composed of two ensembles of ${}^{171}\mathrm{Yb}$ atoms sequentially transported into an optical cavity, in connection with the experiment at FEMTO-ST discussed in \autoref{sec:femto}. The system is treated using the second-order cumulant expansion applied to the open quantum dynamics of atoms and cavity.
Starting from the one-class case of identical atoms equally coupled to the cavity, we confirmed and defined the expected thresholds for superradiant emission and the impact of the atom number and the repumping rate on the output power and the spectral linewidth. We then focused the rest of our analysis within these threshold. Including inhomogeneous broadening and variations in the atom–cavity coupling strengths, we showed that the system is robust with respect to such perturbations. Concerning the detunings, this is due to a synchronization effect among atoms with different frequencies for certain values of the repumping rate. Concerning the coupling strengths,  we confirmed that the properties of the superradiant emission can well be reproduced using a single-class approach with a specific effective coupling strength. We then moved to the two-site configuration specific to the FEMTO-ST setup, using for each site equally detuned atoms, being the two ensembles characterized by a different detuning. For balanced atom numbers, synchronization of atoms among the two sites produces a single spectral line with a narrow linewidth. This is also the case for imbalanced atom numbers, with the central frequency of the spectrum following the weighted central frequency of the system. Such results show that the sequential transport scheme can in principle support continuous superradiant emission for metrological applications, provided that the relative frequency of the atoms in the two sites is controlled at the level required for the targeted frequency stability.

The current analysis is restricted to steady-state configurations with fixed atom numbers in each ensemble. Such configurations can be viewed as successive operating points of the sequential transport process, assuming that atomic decay from, and injection into, the cavity occur on timescales slow compared to the time required to reach the superradiant steady state. Under this separation of timescales, the steady-state spectral characteristics analyzed here are expected to reflect the real time dynamics of the system. From the perspective of the master equation, rather than performing a full time integration including injection and decay processes, we analyzed a sequence of steady-state configurations in which the atom number $N$ is updated. The characteristics of the system such as the cavity field spectrum, linewidth, and power can be evaluated for each configuration and mapped onto the real time operation, providing a practical approach to asses the behavior of the sequential loading strategy. A more detailed investigation going beyond this approximation represents a natural next step and an important direction for future work.

In addition, a more comprehensive model accounting for the four Zeeman sublevels of ${}^{171}\mathrm{Yb}$ would be also of significant interest. In particular, when the magnetic-field-induced frequency shifts become comparable to or smaller than the repumping rate, the emission from different Zeeman sublevels may become synchronized. This synchronization could produce a signal that is naturally insensitive to magnetic fields, providing insight into the robustness of superradiant lasers for metrological applications. 
Moreover, it would be interesting to address analytically the case of at least two classes, aiming to obtain analytic insights about the richness of the occurrence of synchronization in this kind of systems (see for example the behavior found at Figure~\ref{fig:2classes_P_and_nu} where by increasing $R$ we can alternatively obtain one or more peaks in the spectrum).

\section*{Acknowledgments} This project has been supported by the EIPHI Graduate school (contract ANR-17-EURE-0002) and
by the Bourgogne-Franche-Comté Region, by the ANR (project CONSULA, ANR-21-CE47- 0006-02), by First-TF Labex (contract ANR-10-LABX-48-01), and by Oscillator-IMP Equipex (contract ANR-11-EQPX-0033). The authors would like to thank Blandine Guichardaz for the realization of Figure~\ref{fig:femto_exp} and the IT team of the Institut UTINAM for its technical support. Computations have been performed on computers of the Institut Utinam of the Université Marie et Louis Pasteur, supported by the Région Bourgogne-Franche-Comté and the Institut des Sciences de l'Univers (INSU).

%\appendix 
\begin{appendices}
%\textsc{Appendices.}

%\counterwithin*{equation}{section}
%\renewcommand\theequation{\thesection.\arabic{equation}}
\renewcommand{\thesection}{\Alph{section}}% Adjust section printing (from here onward)

\section{One-class analytical considerations: steady-state equations}\label{app:ss}

In this appendix and in the following ones, we provide some analytical considerations concerning the one-class case, when all the atoms can be treated as identical units.

Using Eqs.~\eqref{eqs:Steady_state} in the one-class case, the steady-state intracavity photon number $\braket{a^{\dagger} a}^{\mathrm{st}}$ can be obtained as the positive root of a quadratic equation of the form, $\mathrm{a} \braket{a^{\dagger} a}^{\mathrm{st}~2}+\mathrm{b}\braket{a^{\dagger} a}^{\mathrm{st}} +\mathrm{c}=0 $, where 
\begin{equation}
	\begin{aligned}
		\mathrm{a}&=\frac{2 g \kappa  (N (\gamma +\kappa +T_2^{-1} +R)-\kappa )}{N^2 (\gamma +R) (\gamma +T_2^{-1} +R)},\\
		\mathrm{b}&=\frac{ g \left[\kappa  (T_2^{-1} +2 R)-N (R-\gamma ) (\gamma +\kappa +T_2^{-1} +R)\right]}{ N(\gamma +R) (\gamma +T_2^{-1} +R)}\\
		&+\frac{\kappa}{ 4 g N} \left(\gamma +\kappa +T_2^{-1} +\xi +R+\frac{4 \Delta^2}{\gamma +\kappa +T_2^{-1} +\xi+R}\right),\\
		\mathrm{c}&=-\frac{g R}{\gamma +R}.
	\end{aligned}
\end{equation}
The expression of $\braket{a^{\dagger} a}^{\mathrm{st}}$ in terms of the physical parameters is cumbersome and we do not report it here. From Eqs.~\eqref{eqs:Steady_state}, we obtain for the steady-state values of the other variables:
\begin{equation}\label{Steady state relations}
	\begin{aligned}
		\braket{\sigma^z}^{\mathrm{st}} &= 2\braket{\sigma^+_i\sigma^-_i}^{\mathrm{st}} - 1  =\frac{R-\gamma }{\gamma +R} -\frac{2 \braket{a^{\dagger} a}^{\mathrm{st}} \kappa }{(\gamma +R)N},
		\mathbb{R}\left[\braket{a\sigma^{+}}^\mathrm{st}\right] = -\frac{\braket{a^{\dagger} a}^{\mathrm{st}} \Delta \kappa }{g N (\gamma +\kappa +T_2^{-1} +\xi +R)},\\\mathbb{I}\left[\braket{a\sigma^{+}}^{\mathrm{st}}\right] & =-\frac{\kappa\braket{a^{\dagger} a}^{\mathrm{st}}  }{2 g N},
		\qquad
		\braket{\sigma_{i}^{+} \sigma_{j}^{-}}^{\mathrm{st}} =\frac{\kappa \braket{ a^{\dagger} a}^{\mathrm{st}} }{N (\gamma +R) (\gamma +T_2^{-1} +R)}\left[  R-\gamma -\frac{2 \braket{a^{\dagger} a}^{\mathrm{st}} \kappa}{N}\right].
	\end{aligned}
\end{equation}
We notice that
\begin{equation}\label{app:sign_change}
	\braket{\sigma_{i}^{+} \sigma_{j}^{-}}^{\mathrm{st}} =\frac{\kappa\braket{a^{\dagger} a}^{\mathrm{st}} }{N  (\gamma +T_2^{-1} +R)}\braket{\sigma^z}^{\mathrm{st}},
\end{equation}
showing that the sign of $\braket{\sigma_{i}^{+} \sigma_{j}^{-}}^{\mathrm{st}}$ is equal to the sign of $\braket{\sigma^z}^{\mathrm{st}}$.

\section{Simplified expressions: large \texorpdfstring{$\kappa$ and $R$}{kappa and R} limit}\label{app:simple}

In our experimental setup, the cavity decay rate $\kappa$ is much larger than the other rates in the system, such as $\gamma$, $T_2^{-1}$, and $\xi$. The same holds for $R$, though experimentally its value can be varied over a wide interval, so it is not neglected compared to $\kappa$. In order to reach superradiance, the repumping rate has to be above the lower threshold, i.e., $R>\gamma$. Here we focus on the case $R\gg \{\gamma, T_2^{-1}, \xi \}$ and note that $\Delta$ is also expected to be much smaller than $\kappa$. Since the number of atoms is expected to be very large, we can use the approximations $N-1 \approx N$ and $N-2 \approx N$. With such conditions, the equation governing the steady-state value $\braket{a^{\dagger} a}^{\mathrm{st}}$ is given by the positive solution of a second order equation, $\mathrm{\tilde{a}} \braket{a^{\dagger} a}^{\mathrm{st}~2}+\mathrm{\tilde{b}}\braket{a^{\dagger} a}^{\mathrm{st}} +\mathrm{\tilde{c}}=0 $, where (using $N-2\approx N$)
\begin{equation}
	\begin{aligned}
		\mathrm{\tilde{a}}&=\frac{2 g \kappa  (\kappa +R)}{N R^2},\quad
		\mathrm{\tilde{b}}=(\kappa +R)\left(\frac{\kappa}{4 g  N}-\frac{  g}{R}\right) ,\quad
		\mathrm{\tilde{c}}=-g.
	\end{aligned}
\end{equation}
In this limit we then get the following steady-state values of the variables, considering the condition $N\ge R \kappa/(4 g^2)$ in which we first obtain $\braket{a^{\dagger} a}^{\mathrm{st}}$ and using $N\gg8$:
\begin{equation}\label{Steady state relations simple large N}
	\begin{aligned}
		\braket{a^{\dagger} a}^{\mathrm{st}} & \approx \frac{N R}{2 \kappa}-\frac{R^2}{8 g^2},\qquad 
		\braket{\sigma^z}^{\mathrm{st}} \approx 1 -\frac{2 \braket{a^{\dagger} a}^{\mathrm{st}} \kappa }{R N}\approx\frac{R \kappa}{4 g^ 2 N},
		\\
		\mathbb{R}[\braket{a\sigma^{+}}^{\mathrm{st}}]& \approx -\frac{\braket{a^{\dagger} a}^{\mathrm{st}} \Delta \kappa }{g N (\kappa  +R)}\approx \frac{\Delta R \left(\kappa  R-4 g^2 N\right)}{8 g^3 N (\kappa +R)},
		\qquad 
		\mathbb{I}[\braket{a\sigma^{+}}^{\mathrm{st}}]  =-\frac{\kappa\braket{a^{\dagger} a}^{\mathrm{st}}  }{2 g N}\approx \frac{R \left(\kappa  R-4 g^2 N\right)}{16 g^3 N},
		\\
		\braket{\sigma_{i}^{+} \sigma_{j}^{-}}^{\mathrm{st}}& \approx\frac{\braket{a^{\dagger} a}^{\mathrm{st}} }{N R^2}\left[\kappa  R-\frac{2 \braket{a^{\dagger} a}^{\mathrm{st}} \kappa ^2}{N}\right]\approx \frac{\kappa  R \left(4 g^2 N-\kappa  R\right)}{32 g^4 N^2}.
	\end{aligned}
\end{equation}
Several considerations can be derived from the above equations. For the atom–atom correlations $\braket{\sigma_{i}^{+} \sigma_{j}^{-}}^{\mathrm{st}}$, one finds that they reach a maximum at $N=\kappa R/(2 g^2)$ (equivalently, for fixed $N$, at $R=2 g^2 N/\kappa$), where the value of the maximum is $1/8$. At this particular value of $N$, we have $\braket{a^{\dagger} a}^{\mathrm{st}}\approx g^2 N^2/(2 \kappa^2)$ (this is a maximum with respect to $R$ at $R=2 g^2 N/\kappa$), $\braket{\sigma^z}^{\mathrm{st}}\approx1/2$, $\mathbb{R}[\braket{a\sigma^{+}}^{\mathrm{st}}]\approx-\frac{\Delta g N}{2 (\kappa^2+2 g^2 N)}$, and $\mathbb{I}[\braket{a\sigma^{+}}^{\mathrm{st}}]\approx-g N/(4 \kappa )$ (this is a minimum with respect to $R$ at $R=2 g^2 N/\kappa$). As for the threshold $R_{\mathrm{min}}=\gamma$, Eq.~\eqref{Steady state relations} shows that for $R<\gamma$ we have $\braket{\sigma_{i}^{+} \sigma_{j}^{-}}^{\mathrm{st}}<0$. In the above equations for the limit considered here the steady-state correlations are positive, $\braket{\sigma_{i}^{+} \sigma_{j}^{-}}^{\mathrm{st}}>0$. Using the equations derived for the large $N$ limit in the next subsection [Eqs.~\eqref{SpmijN}], it follows that for a large enough value of $N$ one has $\braket{\sigma_{i}^{+} \sigma_{j}^{-}}^{\mathrm{st}}>0$ for $R>\gamma$ provided that $\kappa R>4 g^2$ (provided the system is in the same limit considered in this subsection).

When $N \leq R\kappa /(4 g^2)$ and at the same time $N \gg 8$, we have
\begin{equation}\label{Steady state relations small N}
	\begin{aligned}
		\braket{a^{\dagger} a}^{\mathrm{st}} & \approx 0,\quad
		\braket{\sigma^z}^{\mathrm{st}}  \approx 1,
		\quad
		\mathbb{R}[\braket{a\sigma^{+}}^{\mathrm{st}}] \approx 0,
		\quad \mathbb{I}[\braket{a\sigma^{+}}^{\mathrm{st}}]  \approx 0,
		\quad
		\braket{\sigma_{i}^{+} \sigma_{j}^{-}}^{\mathrm{st}}\approx 0.
	\end{aligned}
\end{equation}

\section{Simplified expressions: large \texorpdfstring{$N$}{N} limit}\label{app:largeN}

The steady solutions in Eqs.~\eqref{Steady state relations} can be used to compute the expression of $\braket{a^{\dagger} a}^{\mathrm{st}}$ and the other variables in the limit of a very large number of atoms $N$:
\begin{equation}\label{ada large N}
	\begin{aligned}
		\braket{a^{\dagger} a}^{\mathrm{st}}_{N\rightarrow \infty} & = \kappa  R (\gamma +T_2^{-1} +R) \\
		&\times  \frac{\left[4 T_2^{-1}  (\gamma +R)+16 \gamma  R -\frac{(R-\gamma ) (\gamma +R) (\gamma +T_2^{-1} +R) \left(4 \Delta^2+(\gamma +\kappa +T_2^{-1} +\xi +R)^2\right)}{g^2 (\gamma +\kappa +T_2^{-1} +\xi +R)}\right]}{4 N (R-\gamma )^3 (\gamma +\kappa +T_2^{-1} +R)^2}\\
		&+\frac{R (\gamma +T_2^{-1} +R)}{(\gamma -R) (\gamma +\kappa +T_2^{-1} +R)},\qquad &\mathrm{for} R< \gamma\\
		& =\sqrt{\frac{ N \gamma   (2 \gamma +T_2^{-1} )}{2  \kappa   (2 \gamma +\kappa +T_2^{-1} )}} , \qquad &\mathrm{for} R= \gamma 
		\\
		& = \frac{N (R-\gamma )}{2 \kappa } ,\qquad &\mathrm{for} R> \gamma
	\end{aligned}
\end{equation}

\begin{equation} \label{sigmaz large N}
	\begin{aligned}
		\braket{\sigma^z}^\mathrm{st}_{N\rightarrow \infty}  =& \frac{R-\gamma }{\gamma +R}  
		+\frac{2 \kappa  R (\gamma +T_2^{-1} +R)}{N (R-\gamma ) (\gamma +R) (\gamma +\kappa +T_2^{-1} +R)}, \qquad &\mathrm{for} \quad  R< \gamma
		\\
		=&\sqrt{\frac{\kappa  (2 \gamma +T_2^{-1} )}{2 \gamma  N (2 \gamma +\kappa +T_2^{-1} )}} , \qquad &\mathrm{for} \quad R= \gamma 
		\\
		= & \kappa  (\gamma +T_2^{-1} +R) \\ 
		&\times \frac{\left[(\gamma +\kappa +T_2^{-1} +\xi +R)^2+4 \Delta^2 -\frac{4 g^2 (\gamma +\kappa +T_2^{-1} +\xi +R)}{R-\gamma }\right]}{4 g^2 N (\gamma +\kappa +T_2^{-1} +R) (\gamma +\kappa +T_2^{-1} +\xi +R)} ,\qquad &\mathrm{for} \quad R> \gamma
	\end{aligned}
\end{equation}

\begin{equation}
	\begin{aligned}
		\mathbb{R}\left[\braket{a\sigma^{+}}^{\mathrm{st}}_{ N\rightarrow \infty}\right]  =&\frac{\Delta \kappa  R (\gamma +T_2^{-1} +R)}{g N (R-\gamma ) (\gamma +\kappa +T_2^{-1} +R) (\gamma +\kappa +T_2^{-1} +\xi +R)}, \hspace{1.9cm} \mathrm{for} \quad R< \gamma\\
		=& -\frac{(\Delta \kappa ) }{g (2 \gamma +\kappa +T_2^{-1} +\xi )} \sqrt{\frac{\gamma  (2 \gamma +T_2^{-1} )}{2 \kappa  N (2 \gamma +\kappa +T_2^{-1} )}} ,\hspace{3.05cm} \mathrm{for} \quad R= \gamma 
		\\
		=& \Delta \kappa  (\gamma +R) (\gamma +T_2^{-1} +R) \\
		&\times \frac{(R-\gamma ) \left(4 \Delta^2+(\gamma +\kappa +T_2^{-1} +\xi +R)^2\right)-4 g^2 (\gamma +\kappa +T_2^{-1} +\xi +R)}{8 g^3 N (R-\gamma ) (\gamma +\kappa +T_2^{-1} +R) (\gamma +\kappa +T_2^{-1} +\xi +R)^2}\\ &  +\frac{\Delta (R-\gamma )}{2 g (\gamma +\kappa +T_2^{-1} +\xi +R)} ,  \hspace{6.0cm} \mathrm{for}\quad R< \gamma
	\end{aligned}
\end{equation}

\begin{equation}\label{SpmijN}
	\begin{aligned}
		\mathbb{I}\left[ \braket{a\sigma^{+}}^\mathrm{st}_{ N\rightarrow \infty} \right] =&\frac{\kappa  R (\gamma +T_2^{-1} +R)}{2 g N (R-\gamma ) (\gamma +\kappa +T_2^{-1} +R)}, &\mathrm{for} \: R< \gamma\\
		= &-\frac{1}{2 g}\sqrt{\frac{\gamma  \kappa  (2 \gamma +T_2^{-1} )}{2 N (2 \gamma +\kappa +T_2^{-1} )}},  &\mathrm{for} \: R= \gamma 
		\\
		=&-\kappa  (\gamma +R) (\gamma +T_2^{-1} +R)\\
		& \times \frac{\frac{4 g^2 (\gamma +\kappa +T_2^{-1} +\xi +R)}{R-\gamma }-4 \Delta^2-(\gamma +\kappa +T_2^{-1} +\xi +R)^2}{16 g^3 N (\gamma +\kappa +T_2^{-1} +R) (\gamma +\kappa +T_2^{-1} +\xi +R)}-\frac{R-\gamma }{4 g},& \mathrm{for} \: R> \gamma
	\end{aligned}
\end{equation}

\begin{equation}
	\begin{aligned}
		\braket{\sigma_{i}^{+} \sigma_{j}^{-}}^\mathrm{st}_{   N\rightarrow \infty}  = & -\frac{\kappa  R}{N (\gamma +R) (\gamma +\kappa +T_2^{-1} +R)} ,\qquad &\mathrm{for} \quad R< \gamma\\
		= &-\frac{\kappa }{2 N (2 \gamma +\kappa +T_2^{-1} )}, \qquad &\mathrm{for} \quad R= \gamma 
		\\
		= & \frac{\kappa (R-\gamma ) \left(4 \Delta^2+(\gamma +\kappa +T_2^{-1} +\xi +R)^2\right)}{8 g^2 N (\gamma +\kappa +T_2^{-1} +R) (\gamma +\kappa +T_2^{-1} +\xi +R)} \\
		& -\frac{\kappa (4 g^2) (\gamma +\kappa +T_2^{-1} +\xi +R)}{8 g^2 N (\gamma +\kappa +T_2^{-1} +R) (\gamma +\kappa +T_2^{-1} +\xi +R)}.\qquad &\mathrm{for} \quad R> \gamma
	\end{aligned}
\end{equation}

In the limit of large $N$, the above expressions provide useful predictions for experimental setups which operate in this regime. For example, since the output power of the laser is proportional to $\braket{a^{\dagger} a}^{\mathrm{st}}$, it scales linearly with the number of atoms $N$ when $R>\gamma$. In the same range of $R>\gamma$, we have that $\braket{\sigma_z}^{\mathrm{st}}$ is expected to go to zero (corresponds to equally populated ground and excited atomic states), the atom-field correlations $\braket{a\sigma^{+}}^{\mathrm{st}}$ go to a constant value, and the atom-atom correlations $\braket{\sigma_{i}^{+} \sigma_{j}^{-}}^{\mathrm{st}}$ go to zero.
\begin{figure}[h!]
	\centering
	\includegraphics[width=0.7\linewidth]{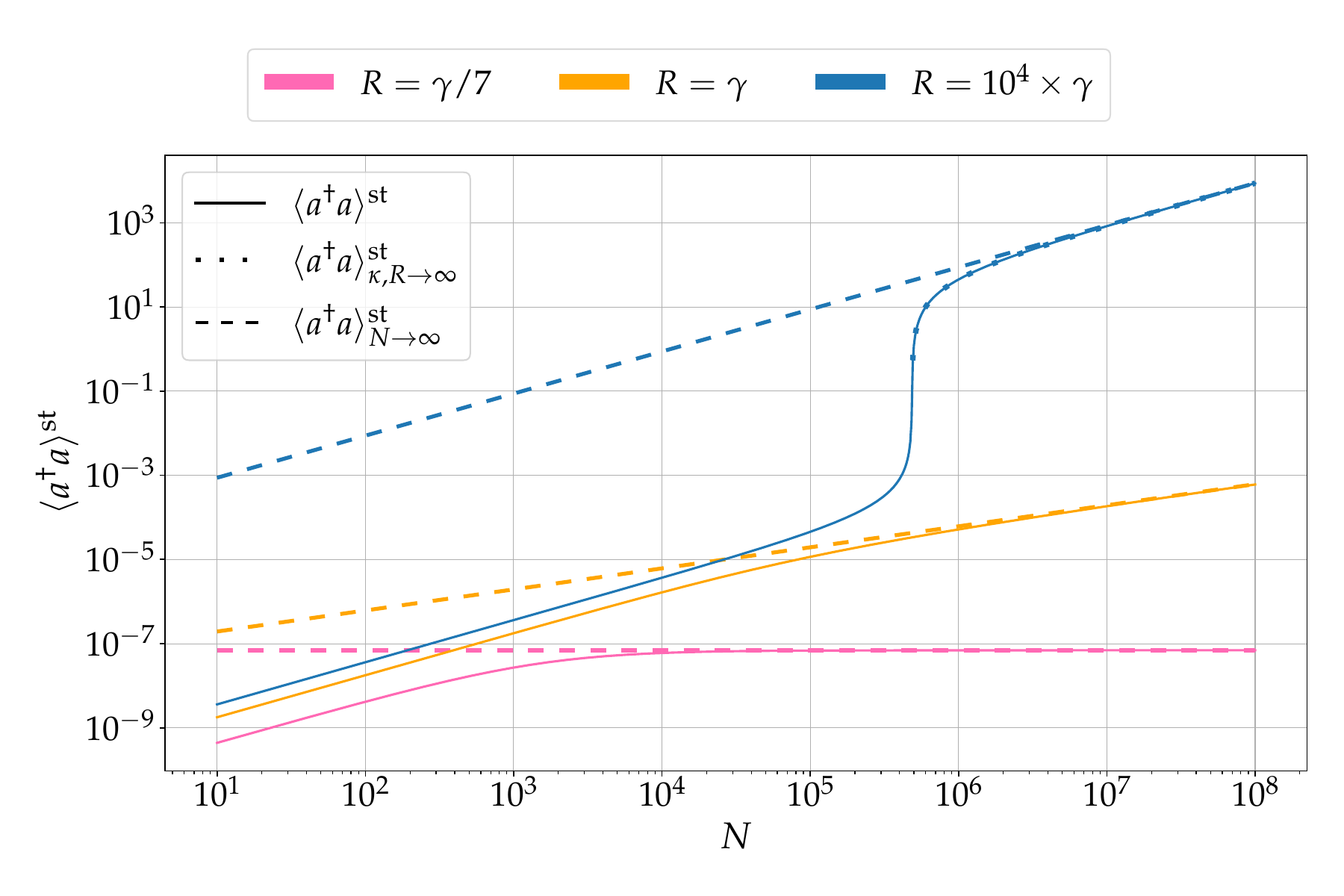}
	\caption{Number of intracavity photons at the steady-state $\braket{a^{\dagger}a}^{\mathrm{st}}$  as a function of atom number $N$ for various values of $R$, for  $\gamma = 2\pi \times 7~\mathrm{mHz} $, $\kappa = 2\pi \times 400 ~\mathrm{kHz} $, $T_2^{-1} = 1 ~\mathrm{rad.s^{-1}}$, $g = 2\pi\times 3.8~ \mathrm{Hz}$, $\Delta = 0$, and $\xi = 0$. }
	\label{fig:steady variables N}
\end{figure}

In Figure~\ref{fig:steady variables N}, we show $\braket{a^{\dagger}a}^{\mathrm{st}}$ as a function of $N$ for three values of $R$, smaller than, equal to, and larger than $\gamma$. Each case is compared with the corresponding large $N$ limit expressions given in Eqs.~\eqref{ada large N} (for $R<\gamma$ we prefer to consider only the large $N$ constant value $\frac{R (\gamma +T_2^{-1} +R)}{(\gamma -R) (\gamma +\kappa +T_2^{-1} +R)}$), and in the case $R=10^4 \gamma$ we also consider the large $\kappa$ and $R$ expression, given in Eqs.~\eqref{Steady state relations simple large N}. This is valid for $N\ge R \kappa/(4 g^2)\approx 4.85\times 10^5$. For smaller values of $N$, in this limit we find $\braket{a^{\dagger}a}^{\mathrm{st}}\approx 0$ [Eqs.~\eqref{Steady state relations small N}], which cannot be reported in the log-log scale used in the figure. The results show that for large $N$, $\braket{a^{\dagger}a}^{\mathrm{st}}$ approaches a constant value for $R<\gamma$, increases as $\sqrt{N}$ for $R=\gamma$, and increases linearly in $N$ for $R>\gamma$.
The last case $R>\gamma$, is the one in which the superradiant laser operates and then, for any fixed value of $R$, the power of the laser is expected to become proportional to $N$ for values of this variable large enough to have access to this limit.
\begin{figure}[h!]
	\centering
	\includegraphics[width=0.7\linewidth]{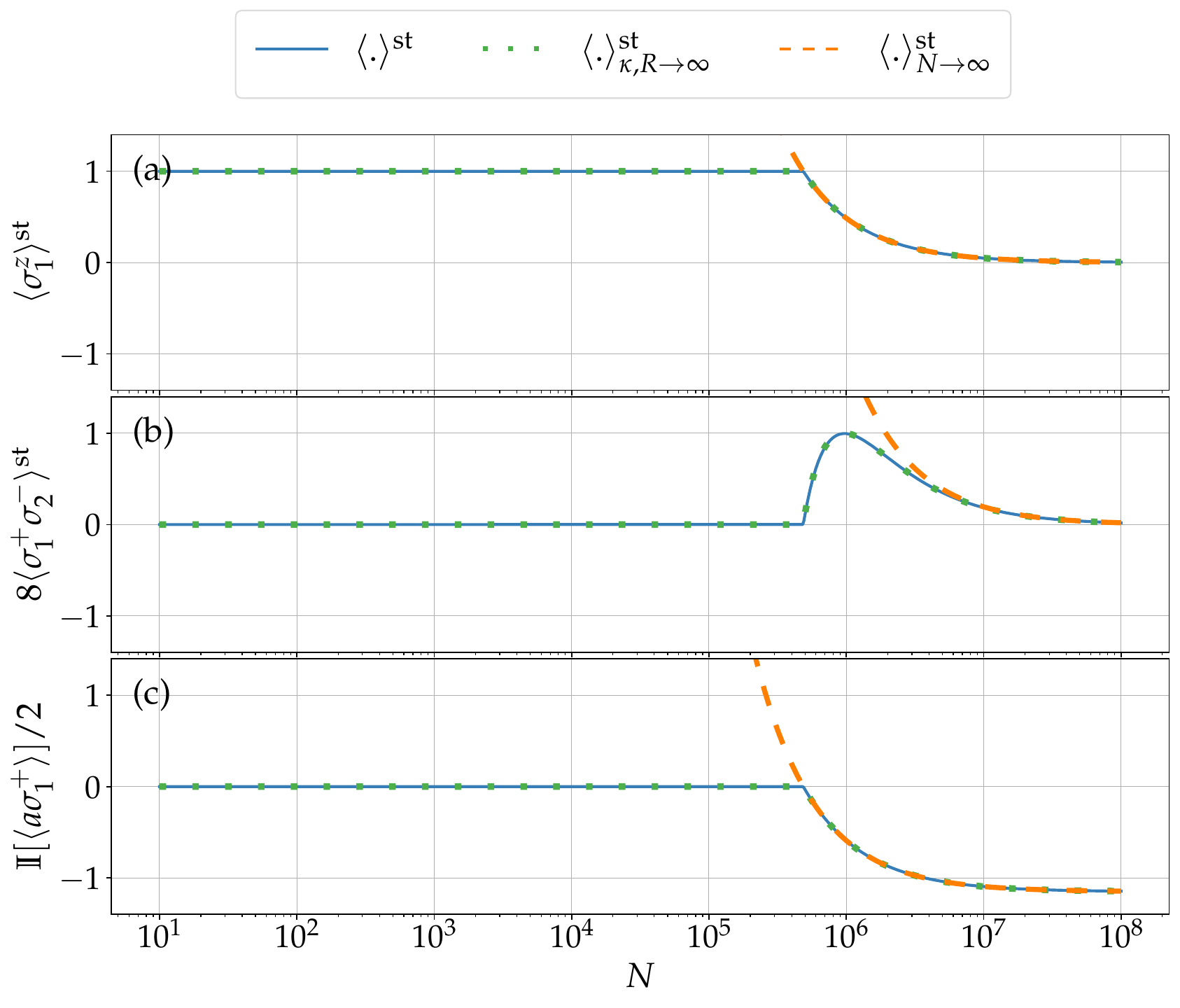}
	\caption{Atomic variables  as a function of  $N$ for  $R=10^4 \gamma$, for  $\gamma = 2\pi \times 7~\mathrm{mHz} $, $\kappa = 2\pi \times 400 ~\mathrm{kHz} $, $T_2^{-1} = 1 ~\mathrm{rad.s^{-1}}$, $g = 2\pi\times 3.8~\mathrm{Hz}$, $\Delta = 0$, and $\xi = 0$. }
	\label{fig:atomic steady variables N}
\end{figure}

In Figure\ref{fig:atomic steady variables N}, we show for the case $\Delta=\xi=0$, the steady-state variables which involve the atoms, $\braket{\sigma^{z}}^{\mathrm{st}}$, $\braket{\sigma_{i}^{+} \sigma_{j}^{-}}^{\mathrm{st}}$, and $\mathbb{I}\left[\braket{a \sigma^{+}}^{\mathrm{st}}\right]$ (the real part is zero for $\Delta=0$) as a function of $N$ for $R=10^4 \gamma$. This is compared with both the large $\kappa$ and $R$ expressions given in Eqs.~\eqref{Steady state relations small N} for $N< R\kappa/(4 g^2)\equiv N^* \approx 4.85 \times 10^5$ and in Eqs.~\eqref{Steady state relations simple large N} for larger values of $N$, and the large $N$ limit expressions given in Eqs.~\eqref{ada large N}. We can observe how  the large $\kappa$ and $R$ expressions work well for any $N$ while the large $N$ expressions are valid above specific values of $N$. The variable $\braket{\sigma^{z}}^{\mathrm{st}}$ here starts from 1 for $N< N^*$ and decreases to zero for larger values of $N$. Both $\braket{\sigma_{i}^{+} \sigma_{j}^{-}}^{\mathrm{st}}$ and $\mathbb{I}\left[\braket{a \sigma^{+}}^{\mathrm{st}}\right]$ start from zero for $N< N^*$. For larger values of $N$, $\braket{\sigma_{i}^{+} \sigma_{j}^{-}}^{\mathrm{st}}$ increases up to a maximum of $1/8$ for $N =2 N^* $ and then decreases back to zero. For this specific value of $N$, we have $\braket{\sigma^{z}}^{\mathrm{st}}\approx 1/2$, $\braket{\sigma_{i}^{+} \sigma_{j}^{-}}^{\mathrm{st}}\approx 1/8$, and $\mathbb{I}\left[\braket{a \sigma^{+}}^{\mathrm{st}}\right]\approx -R/(8 g) \approx -2.30$. For values of $N>N^*$, $\mathbb{I}\left[\braket{a \sigma^{+}}^{\mathrm{st}}\right]$ decreases towards a constant minimum value approximately equal to  $-R/(4 g) \approx -4.61$. This analysis is based on the approximated expressions that are valid in the large $\kappa$ and $R$ limit.

\section{Cavity field spectrum}\label{app:spectrum}

In the case when all the atoms have the same parameters, the $N+1$ equations in Eqs.~\eqref{eq:time_evo} reduce to two equations:
\begin{equation}\label{app:2eqs:matrix}
	\frac{d}{d\tau} 
	\begin{pmatrix}
		\braket{a^\dagger (\tau+t) a(t)} \\
		\braket{\sigma^\dagger(\tau+t) a(t)} 
	\end{pmatrix} 
	= 
	\begin{pmatrix} 
		c_1 & \mathrm{i}c_2 \\ 
		\mathrm{i}c_3 & c_4 
	\end{pmatrix} 
	\begin{pmatrix}
		\braket{a^\dagger (\tau+t) a(t)} \\
		\braket{\sigma^\dagger(\tau+t) a(t)} 
	\end{pmatrix},
\end{equation}
where the coefficients are defined as $ c_1 = -(\kappa+\xi)/2$, $c_2 = gN$, $c_3 = -g\braket{\sigma^z}^\mathrm{st}$, and $c_4 = c_{4\mathbb{R}}+\mathrm{i} c_{4\mathbb{I}} = -(\gamma + R + 1/T_2)/2 - \mathrm{i}\Delta$. The system can be solved by applying the Laplace transform, using the standard operational rule $\mathcal{L}\left[\frac{d}{d \tau} f(\tau)\right]=s F(s)-$ $f(0)$, with the initial conditions $\braket{a^{\dagger}(0) a(0)} = \braket{a^{\dagger}a}^{\mathrm{st}}$ and $\braket{\sigma^{+}(0) a(0)} = \braket{\sigma^{+}a}^{\mathrm{st}}$. Solving the resulting set of equations gives
\begin{equation}\label{Spectrum1}
	\braket{a^{\dagger}(s) a(0)} = \frac{(s - c_4)\braket{a^{\dagger} a}^{\mathrm{st}} + c_2\braket{\sigma^{+}a}^{\mathrm{st}}}{(s-c_1)(s-c_4) - c_2c_3}.
\end{equation}
Using Eqs.~\eqref{Steady state relations}, the above expression leads to
\begin{equation}\label{app:ata:spectrum:N}
	\braket{a^{\dagger}(\omega) a(0)} =\braket{a^{\dagger} a}^{\mathrm{st}}\frac{\frac{1}{2} (\gamma +\kappa +T_2^{-1} +R)+i \left(\frac{\Delta  (\gamma +T_2^{-1} +\xi +R)}{\gamma +\kappa +T_2^{-1} +\xi +R}+\omega \right) }{\left(\frac{\kappa +\xi }{2}+i \omega \right) \left[\frac{1}{2} (\gamma +T_2^{-1} +R)+i (\Delta +\omega ) \right]-g^2 N \braket{\sigma^z}^{\mathrm{st}}},
\end{equation}
where we used $s=\mathrm{i} \omega$ to correctly compute $S(\omega)$, which is given by $ 2 \mathbb{R}\biggl\{ \braket{a^{\dagger}(\omega) a(0)} \biggr\}$. 

The above expression shows that in the large $N$ limit, where $\braket{a^{\dagger} a}^{\mathrm{st}}$ increases linearly with $N$, as shown in Eqs.~\eqref{ada large N}, and $\braket{\sigma^z}^{\mathrm{st}}$ scales inversely proportional to $N$, as shown in Eqs.~\eqref{sigmaz large N}, the spectrum takes the form of a function of $\omega$ multiplied by $N$. It follows that in this limit the width of the spectrum becomes independent of $N$. This behavior is observed in Figure~\ref{fig:power_and_linewidth_1class}(b) for large values of $N$ and for values of $R$ larger than $\gamma$ up to some value (for $R$ very large we can expect that the limit of large $N$ occurs for values of $N$ larger than the ones considered in this figure). The independence of the linewidth from $N$ is reflected by the vertical behavior of the white contour lines at constant values of $R$.

We note that the linewidth stays independent of $N$ also in the case considered in \autoref{app:simple} for $N\ge R \kappa/(4 g^2)$, since $\braket{\sigma^z}^{\mathrm{st}}$ is inversely proportional to $N$, as shown in Eqs.~\eqref{Steady state relations simple large N}.

From Eq.~\eqref{app:ata:spectrum:N}, the spectrum can be put under the final form
\begin{equation}\label{eq:Spectrum}
	\begin{aligned}
		& S(\omega)=\braket{a^{\dagger} a}^{\mathrm{st}}\Bigg[(\gamma +\kappa +T_2^{-1} +R) \\ & \times \left( \frac{(\kappa +\xi ) \left(4 \Delta ^2+(\gamma +T_2^{-1} +R) (\gamma +\kappa +T_2^{-1} +\xi +R)\right)}{4 (\gamma +\kappa +T_2^{-1} +\xi +R)}-g^2 N \braket{\sigma^z}^{\mathrm{st}} \right)+2 \Delta  \xi  \omega   +\xi  \omega^2\Bigg]\Bigg/ \\&  \Bigg[\frac{(\kappa +\xi )^2}{16}  \left(4 \Delta ^2+(\gamma +T_2^{-1} +R)^2\right)-\frac{1}{2} g^2  (\kappa +\xi ) (\gamma +T_2^{-1} +R)N \braket{\sigma^z}^{\mathrm{st}} +g^4 N^2 \braket{\sigma^z}^{\mathrm{st}~2}\\&   +\Delta \left(\frac{(\kappa +\xi )^2}{2}   +2  g^2 N \braket{\sigma^z}^{\mathrm{st}}\right)\omega + \left(\frac{1}{4} \left(4 \Delta ^2+(\kappa +\xi )^2+(\gamma +T_2^{-1} +R)^2\right)+2 g^2 N \braket{\sigma^z}^{\mathrm{st}}\right) \\
		&  \times \omega ^2 +2 \Delta  \omega ^3 +\omega ^4 \
		\Bigg] .   
	\end{aligned}
\end{equation}

An alternative way to manipulate Eq.~\eqref{Spectrum1} consists in transforming it into
\begin{equation}
	\braket{a^{\dagger}(s) a(0)} = \frac{(s - c_4)\braket{a^{\dagger} a}^{\mathrm{st}} + c_2\braket{\sigma^{+}a}^{\mathrm{st}}}{(s - \lambda_+)(s - \lambda_-)},
\end{equation}
where
\begin{equation}\label{app:lambda_pm}
	\lambda_{\pm}=\frac{1}{2} \left((c_1+c_4 ) \pm  \sqrt{\left(c_1+c_4 \right)^2-4\left(c_1 c_4  -c_2 c_3\right)}\right) = a_{\pm} + \mathrm{i} b_{\pm}.
\end{equation}
By using the method of partial fractions, $\braket{a^{\dagger}(s) a(0)}$ can be then expressed as
\begin{equation}\label{Partial fractions}
	\braket{a^{\dagger}(s) a(0)} = \frac{X}{(s - \lambda_+)} + \frac{Y}{(s - \lambda_-)},
\end{equation}
where
\begin{equation}
	X = \frac{c_4 \braket{a^{\dagger} a}^{\mathrm{st}} - c_2 \braket{\sigma_A^{+} a}^{\mathrm{st}} - \lambda_{+} \braket{a^{\dagger} a}^{\mathrm{st}}}{\lambda_{-} - \lambda_{+}}, \quad Y = \frac{-c_4 \braket{a^{\dagger} a}^{\mathrm{st}} + c_2 \braket{\sigma_A^{+} a}^{\mathrm{st}} + \lambda_{-} \braket{a^{\dagger} a}^{\mathrm{st}}}{\lambda_{-} - \lambda_{+}} .
\end{equation}
From the above equation we can find that $ X + Y = \braket{a^{\dagger} a}^{\mathrm{st}}$, with $\braket{a^{\dagger} a}^{\mathrm{st}}$ a real value. It then follows that $X  = X_{\mathbb{R}} + \mathrm{i} X_{\mathbb{I}}$ and $Y  = Y_{\mathbb{R}} - \mathrm{i} X_{\mathbb{I}}$.

The inverse Laplace transform of Eq.~\eqref{Partial fractions} then gives
\begin{equation}\label{inverseLT}
	\braket{a^{\dagger}(\tau) a(0)} = Xe^{\lambda_+\tau} + Ye^{\lambda_-\tau}.
\end{equation}
The structure of $\lambda_{\pm}$ determines that the second term in the above Eq.~\eqref{inverseLT} decays to zero faster than the first \cite{PhD_DavidTieri, PhD_SwadheenDubey}.

To obtain the final expression for the spectrum we have to compute $S(\omega) \equiv 2 \mathbb{R}\biggl\{ \braket{a^{\dagger}(\omega) a(0)} \biggr\}$.
From Eq.~\eqref{Partial fractions}, we have
\begin{equation}
	\begin{aligned}
		\braket{a^{\dagger}(\omega) a(0)}  = &\frac{X_{\mathbb{R}} + \mathrm{i} X_{\mathbb{I}}}{-a_+ +i(\omega - b_+)} + \frac{Y_{\mathbb{R}} - \mathrm{i} X_{\mathbb{I}}}{-a_- +i(\omega - b_-)}   \\
		=& \frac{-a_+X_{\mathbb{R}}+X_{\mathbb{I}}(\omega-b_+) - \mathrm{i}a_+X_{\mathbb{I}} -\mathrm{i} X_{\mathbb{R}}(\omega-b_+)}{a_+^2 +(\omega-b_+)^2}\\& + \frac{-a_-Y_{\mathbb{R}}-X_{\mathbb{I}}(\omega-b_-) + \mathrm{i}a_-X_{\mathbb{I}} -\mathrm{i} Y_{\mathbb{R}}(\omega-b_-)}{a_-^2 +(\omega-b_-)^2},
	\end{aligned}
\end{equation}
leading to
\begin{equation}
	S(\omega) = 
	2\frac{-a_+X_\mathbb{R} + X_{\mathbb{I}}(\omega - b_+)}{a_+^2 + (\omega - b_+)^2} 
	+ 2\frac{-a_- Y_\mathbb{R} - X_{\mathbb{I}}(\omega - b_-)}{a_-^2 + (\omega - b_-)^2}.
\end{equation}
For $X_{\mathbb{I}} =0$, we have two Lorentzian functions centered at $b_+$ and $b_-$. In the general case $X_{\mathbb{I}} \neq 0$, each Lorentzian is modified by a linear function at the numerator. 

\subsection*{Cavity field spectrum for \texorpdfstring{$\Delta=\xi=0$}{Delta=xi = 0}}

In the case $\Delta=\xi=0$, Eq.~\eqref{eq:Spectrum} simplifies to 
\begin{equation} \label{app:Spectrum special case}
	\begin{aligned}
		& S(\omega)=\braket{a^{\dagger} a}^{\mathrm{st}} \Bigg[(\gamma +\kappa +T_2^{-1} +R)\left( \frac{\kappa  (\gamma +T_2^{-1} +R) (\gamma +\kappa +T_2^{-1} +R)}{4 (\gamma +\kappa +T_2^{-1}  +R)}-g^2 N \braket{\sigma^z}^{\mathrm{st}} \right)\Bigg]\Bigg/  \\ &
		\Bigg[\frac{\kappa^2}{16} (\gamma +T_2^{-1} +R)^2 -\frac{1}{2} g^2  \kappa  (\gamma +T_2^{-1} +R)N \braket{\sigma^z}^{\mathrm{st}}+g^4 N^2 \braket{\sigma^z}^{\mathrm{st}~2} \\ & + \left(\frac{1}{4} \left(\kappa^2+(\gamma +T_2^{-1} +R)^2\right)+2 g^2 N \braket{\sigma^z}^{\mathrm{st}}\right)\omega ^2+\omega ^4 \Bigg] ,   
	\end{aligned}
\end{equation}

where $\braket{a^{\dagger} a}^{\mathrm{st}}$ and $\braket{\sigma^z}^{\mathrm{st}}$ are the steady-state values for $\Delta=\xi=0$.
The dependence of the above spectrum on $\omega$ has the form 
\begin{equation}\label{app:Spectrum special case form}
	S(\omega)=\frac{X^\prime}{Y^\prime+Z\omega ^2+\omega ^4} ,   
\end{equation}
where the form of $X^\prime$, $Y^\prime$, and $Z$ can be obtained from Eq.~\eqref{app:Spectrum special case}. The spectrum is then symmetric with respect to $\omega=0$ and for $Z>0$, which is valid for $\braket{\sigma^z}^{\mathrm{st}}>0$, condition satisfied in the regime treated in \autoref{app:simple}, $S(\omega)$ has a maximum equal to $X^\prime/Y^\prime$ for $\omega=0$. The central frequency of the spectrum is then equal to zero in the rotating frame of the cavity meaning that it is equal to the cavity frequency. The linewidth $\Delta\nu$ which corresponds to the FWHM of the spectrum, is determined by the equation $S(\omega)=X^\prime/(2Y^\prime)$ leading to
\begin{equation}\label{app:FWHM special case form}
	\Delta\nu=\sqrt{2} \sqrt{\sqrt{4 Y^\prime+Z^2}-Z} .   
\end{equation}
We then find that a possible approach to reduce $\Delta\nu$ is to operate where $Y^\prime\approx 0$ or $Y^\prime\ll Z$.

In the large $\kappa$ and $R$ limit in \autoref{app:simple}, $Y^\prime$ takes the form $\frac{\kappa^2R^2 }{16}   
-\frac{1}{2} g^2  \kappa  R N \braket{\sigma^z}^{\mathrm{st}}+g^4 N^2 \braket{\sigma^z}^{\mathrm{st}~2} $.
For $N\le R \kappa/(4 g^2)$, we have for the atomic variable  $\braket{\sigma^z}^{\mathrm{st}}\approx 1$, which gives $Y^\prime\approx 0$ at $N= R \kappa/(4 g^2)$, and for $N\ge R \kappa/(4 g^2)$, we have $\braket{\sigma^z}^{\mathrm{st}}\approx R \kappa/(4 g^ 2 N)$ and $Y^\prime\approx 0$.

In the large $N$ limit of \autoref{app:largeN}, with $R>\gamma$, we consider the large $\kappa$ and $R$ limit, such that the large $N$ approximation stays valid. For the case $\kappa \gg R$, we find that $Y^\prime=g^4$ and $Z=\kappa^2/4$. This implies that $Y^\prime\ll Z^2$, which allows us to develop Eq.~\eqref{app:FWHM special case form} to obtain
\begin{equation}\label{app:FWHM special case form large N}
	\Delta\nu\approx 2 \sqrt{\frac{Y^\prime}{Z}}=\frac{4 g^2 }{\kappa}.  
\end{equation}
This is equal to the one reported in~\cite{Meiser2009, Kazakov2022}, where the minimal linewidth of a superradiant laser is estimated by $ \frac{4 g^2}{\kappa}$.

\end{appendices}

\bibliographystyle{crunsrt}
\bibliography{references}

\end{document}